\documentclass[namedreferences]{solarphysics}

\usepackage[hyperref,optionalrh,showbiblabels]{spr-sola-addons} 
\usepackage{graphicx}        
\usepackage{color}           
\usepackage{breakurl}        
\usepackage{rotating}
\usepackage{lscape}
\usepackage{enumerate}
\usepackage{tikz}
\usetikzlibrary{decorations.pathreplacing}
\usetikzlibrary{shapes.geometric, arrows}
\tikzstyle{flowbox} = [rectangle, rounded corners, minimum width=7cm, minimum height=1cm,  text width=7cm, text centered, draw=blue]
\tikzstyle{arrow} = [thick,->,>=stealth]

\newcommand{\etal}{{\it et al.}}

\usepackage{multirow}

\hypersetup{pdftitle = The title of my PDF, pdfauthor = My name, pdfsubject= The subject, pdfkeywords = keyword1 keyword2 keyword3} 
\hypersetup{colorlinks = true, linkcolor = red, anchorcolor = red, citecolor = blue, filecolor = red, pagecolor = red, urlcolor = red}

\chardef\us=`\_

\begin{document}

\begin{article}
\begin{opening}

\title{\Large Automated Detection of Accelerating Solar Eruptions using Parabolic Hough Transform\\ {\it Solar Physics}}

\author[addressref={aff1,aff1a,aff1b},corref,email={ritesh.patel@iiap.res.in}]{\inits{R.}\fnm{Ritesh}~\lnm{Patel} \orcid{0000-0001-8504-2725}}
\author[addressref={aff2,aff2a,aff2b},corref,email={vaibhav.pant@iac.es}]{\inits{V.}\fnm{Vaibhav}~\lnm{Pant} \orcid{0000-0002-6954-2276}}
\author[addressref={aff3},corref,]{\inits{P.}\fnm{Priyanka}~\lnm{Iyer}}
\author[addressref={aff1,aff1a,aff4},corref,email={dipu@iiap.res.in}]{\inits{D.}\fnm{Dipankar}~\lnm{Banerjee} \orcid{0000-0003-4653-6823}}
\author[addressref={aff5,aff6},corref]{\inits{M.}\fnm{Marilena}~\lnm{Mierla}\orcid{0000-0003-4105-7364}}
\author[addressref={aff5},corref]{\inits{M.}\fnm{Matthew J.}~\lnm{West}\orcid{0000-0002-0631-2393}}


\address[id=aff1]{Indian Institute of Astrophysics, Koramangala, Bangalore - 560034, India}
\address[id=aff1a]{Aryabhatta Research Institute of Observational Sciences, Nainital 263001, India}
\address[id=aff1b]{University of Calcutta, 87/1, College Street, Kolkata, 700073, India}
\address[id=aff2]{Instituto de Astrofis\'ica de Canarias, E-38205 La Laguna, Tenerife, Spain}
\address[id=aff2a]{Departamento de Astrofis\'ica, Universidad de La Laguna, E-38206 La Laguna, Tenerife, Spain}
\address[id=aff2b]{Department of Mathematics, Centre for mathematical Plasma Astrophysics, KU Leuven, Celestijnenlaan 200B, 3001 Leuven, Belgium}
\address[id=aff3]{University of Delhi, Benito Juarez Marg, South Moti Bagh, New Delhi, 110007, India}
\address[id=aff4]{Center of Excellence in Space Sciences, IISER Kolkata - 741246, India}
\address[id=aff5]{Solar-Terrestrial Center of Excellence, Royal Observatory of Belgium, Avenue Circulaire 3, B-1180 Brussels, Belgium}
\address[id=aff6]{Institute of Geodynamics of the Romanian Academy, Bucharest, Romania}

\runningauthor{{Patel R. \etal}}
\runningtitle{CIISCO}

\begin{abstract}
{  Solar eruptions such as} Coronal Mass Ejections (CMEs) observed in the inner solar corona (up to 4 R$_{\odot}$) show acceleration profiles which appear as parabolic ridges in height-time plots. Inspired by the white-light automated detection algorithms, Computer Aided CME Tracking System (CACTus) and Solar Eruptive Events Detection System (SEEDS), we employ the parabolic Hough Transform for the first time to automatically detect off-disk solar eruptions {from height-time plots.} Due to the limited availability of white-light observations in the inner corona, we use Extreme UltraViolet (EUV) images of the Sun. In this paper we present a new algorithm, CME Identification in Inner Solar Corona (CIISCO), which is based on Fourier motion filtering and the parabolic Hough transform, and demonstrate its implementation using EUV observations taken from  {\it Atmospheric Imaging Assembly} (AIA) on-board the Solar Dynamics Observatory (SDO), {\it Extreme Ultra Violet Imager} (EUVI) on-board the STEREO-A and B satellites, and {\it Sun Watcher using Active Pixel System detector and Image Processing} (SWAP) Imager on-board PROBA2. We show that CIISCO is able to identify any {off-disk} outward moving feature in EUV images. { The use of automated detection algorithms, like CIISCO, can potentially be used to provide early warnings of CMEs if an EUV telescope is located at $\pm$ 90$^{\circ}$ from the Sun-Earth line}, providing CME characteristics and kinematics close to the Sun. This paper also presents the limitations of this algorithm and the prospects for future improvement.
\end{abstract}
\end{opening}

\section{Introduction} \label{sec:intro}

Coronal mass ejections (CMEs) are large scale eruptions of plasma and magnetic field from the solar atmosphere into the interplanetary space, and are most commonly observed in white-light coronagraphs. These eruptions are known to have a three-phase kinematics profile starting with a gradual rise phase followed by an impulsive acceleration phase below 2 R$_{\odot}$, and final phase of constant average speed  \citep{Zhang_2001, Zhang_2004, Bein2011, Majumdar2020ApJM}.
It is now well established that CMEs play an important role in driving space weather \citep{Gosling1993JGR}, and therefore, it is necessary to understand their origin and early development through the { inner (up to 4 R$_\odot$) and outer corona (above 4 R$_\odot$).}
For the last two decades space-based white-light observations of the corona have been made by the {\it Large Angle Spectroscopic COronagraph} (LASCO), which was originally a coronagraph system comprising of three units C1, C2 and C3 capable of observing the Sun from 1.1 to 30 R$_{\odot}$ \citep{Brueckner95}, with the inner coronagraph, C1, having FOV from 1.1 to 3 R$_{\odot}$. More recently, observations have also been made with the {\it Solar TErrestrial RElations Observatory} (STEREO) COR1 coronagraph, which has a FOV extending from 1.4 to 4 R$_{\odot}$ \citep{Howard02}. However, LASCO-C1 stopped observing after 1998 and  STEREO/COR1 images suffer from heavy compression, noise and artefacts. Even with the inner edge of our space-based coronagraphs extending to down to 1.4 R$_\odot$, these imagers struggle to capture the kinematics of eruptions during their acceleration phase. Such observations are important for  understanding the over-arching propagation of an eruption. Models such as Empirical CME Arrival (ECA) \citep{NATG2000, NATG2001} model use initial kinematics of CMEs as an input to predict their arrival times at Earth. A better understanding of their initial properties can help to improve such empirical models. It should be noted that some ground based coronagraphs, such as K-Cor (\citealp{deWijn}), in Mauna Loa Solar Observatory (MLSO), image the lower solar corona (1.05 - 3~R$_{\odot}$). However, ground based imaging has its own set of issues, such as being limited by atmospheric conditions and day time observations. 


Full disk images of the EUV emission corona have been regularly taken over the past two solar cycles, starting from {  \textit{Extreme ultra-violet Imaging telescope}} (EIT) on-board {  \textit{SOlar and Heliospheric Observatory}} (SOHO) \citep{SOHOEIT}, followed by EUVI on-board STEREO \citep{Howard02}, {  \textit{Atmospheric Imaging Assembly}} (AIA) on-board {  \textit{Solar Dynamics Observatory}} (SDO) \citep{AIA}, {  \textit{Sun Watcher using Active Pixel System detector and Image Processing}} (SWAP) on-board {  \textit{PRoject for Onboard Autonomy 2}} (PROBA2) \citep{SWAPinstru, SWAPcalib} and the recent {  \textit{Solar Ultra Violet Imager}} (SUVI) on-board {  \textit{Geostationary Operational Environmental Satellite}} (GOES-R) {  \citep{SUVI2018}}. It has long been known that the coronal emission observed in EUV pass-bands is generated by atomic transitions of different ions present in the solar atmosphere, whereas the white-light corona is observed through photospheric light bouncing off free electrons due to Thomson scattering. As a consequence the coronal features observed in the two types of imager are not always the same. A study of the kinematics of CMEs made by \cite{Bein2011} (and references therein) combining EUVI data from the emission corona with white-light observations from COR-1 and COR-2, suggests that the early phases of CMEs can have acceleration profiles as high as 1000 m s$^{-2}$. With large FOV EUV imagers such as SWAP and EUVI, the eruptive profiles recorded from EUV and white-light observations can be combined to create a more complete picture  of the initial kinematics of a CME. Although an EUV eruption front may not have a one-to-one correspondence with the leading edge of a white-light CME,  it can give us an idea about the over-arching kinematics of propagating eruptions in the inner corona.

The manual detection and tracking of solar eruptions in large datasets is time consuming and subjective. In order to overcome these limitations, algorithms to automatically detect CMEs in coronagraph imagery were developed, starting with CACTus, which uses the linear Hough transform to detect CMEs { as white ridges in height-time plots} of LASCO images \citep{Robbrecht04, LASCOCME}. CACTus was later extended to STEREO/COR2 data. {  It was recently adapted as CACTusCAT\footnote{see \url{sidc.be/cactus/hi/}.} to be used with STRERO/HI1 images \citep{Pant2016}.} Another algorithm,  SEEDS, transforms images from LASCO and the STEREO coronagraphs to polar coordinates in which the intensity is integrated in the radial direction at each position angle (PA) and reduced to one dimensional (1D) arrays. { CMEs are identified in such arrays and  processed to track the leading edge as the outermost boundary of the moving feature in the intensity threshold running difference images.} { A similar approach was also used to automatically detect CMEs in K-Cor images by  \cite{Thompson17}.
In another algorithm, Automatic Recognition of Transient Events and Marseille Inventory from Synoptic maps (ARTEMIS), adaptive filtering and segmentation techniques are used to identify CMEs { as bright streaks} in synoptic maps of LASCO images \citep{Boursier09}. The Coronal Image Processing (CORIMP) algorithm separates quiescent and dynamic coronal structures observed in coronagraph images using deconvolution and { detects CMEs structure using a multi-scale edge detection method \citep{Morgan12, Byrne12} taking in to account CME kinematics and morphology changes}.} {In a recent work \cite{adaptiveBG2019} developed an algorithm based on adaptive background learning to detect CMEs in LASCO/C2 images considering CMEs to be dynamic foreground features in running difference images.}

Recently, algorithms based on machine learning have also been developed for automated CME detection \citep{MLQu2006, MLADABOOST16, ML2017, CAMEL2019}. {On-board algorithms based on an intensity threshold, using running difference polarized brightness coronagraph images, have been developed for  {\it Multi Element Telescope for Imaging and Spectroscopy} (METIS) on-board the Solar Orbiter \citep{Bemporad14}.  Visible Emission Line Coronagraph (VELC)  on-board ADITYA-L1 \citep{VELC17, ADITYA2017} has a simple on-board algorithm for CME detection based on intensity and area threshold \citep{Patel2018, IAUS2018, IAUS2017}.} 


The aforementioned algorithms focus on the white-light data. However, not much work has been done to automatically identify the EUV counterparts of white-light CMEs. Recently, an algorithm was developed to automatically detect and catalog prominence eruptions in SDO/AIA 304 \AA\ \footnote{see \url{https://cdaw.gsfc.nasa.gov/CME_list/autope/}.} observations \citep{Yashiro2020arXiv}. Among the existing automated CME detection methods, CACTus is limited by the use of the linear Hough transform. Since, CMEs in the inner corona accelerate, they may appear as parabolic or higher order ($>2$) polynomial ridges in height-time (r-t) plots. CACTus has been designed to detect straight lines in r-t plots,  the slope of which gives the speed of CMEs, and therefore may miss eruptions or parts of eruptions that are represented as parabolic ridges in the r-t plots of inner corona. A consequence of this will be the algorithm misses important information about the CME acceleration. It should be noted that SEEDS and CORIMP are not successfully implemented on the images taken by STEREO/COR1 that observes the inner corona. 

As discussed above, although there is not an exact correspondence between EUV and white light observations, the structures observed in emission will go some way to providing an approximation of the early characteristics of CMEs observed further out in white light observations. By applying automated detection algorithms to EUV observations of the inner corona we can identify and track the eruptions in their early stages of development. Similar to CACTus, if CIISCO is  applied to a large dataset it can provide a large statistical sample of solar eruption properties in the inner corona.
{ Therefore in this article we present a novel automated method, CMEs Identification in Inner Solar Corona (CIISCO), which identifies the off-disk solar eruptions as intensity enhancements and tracks them as parabolic ridges in height-time plots of EUV images using the parabolic Hough transform.}
This paper is organized as follows. We describe the different datasets used for analysis in Section \ref{sec:method} along with the algorithm. In Section \ref{sec:results}, we present the results of application of the algorithm on several EUV images from different instruments, in particular the large FOV EUV imager SWAP.  We conclude with a summary and discussion in Section \ref{sec:summary}.

\section{Observations and Method of Detection} \label{sec:method}

The erupting plasma leading to solar eruptions can be distinctly seen in both 171/174 \AA\ , and 304 \AA\ EUV pass-bands. The 171/174 \AA\ pass-band EUV imagers observe hotter coronal plasmas, whereas the 304 \AA\ pass-band observes chromospheric temperatures, and hence cooler denser structures such as filaments.
Therefore, we have used full disk EUV images from AIA \citep{AIA}, EUVI \citep{Howard02} and SWAP  \citep{SWAPinstru, SWAPcalib}. The FOV of these instruments are up to 1.3 R$_\odot$, 1.7 R$_\odot$, and 1.7 R$_\odot$ respectively. { The high cadence (12 s) but smaller FOV observations, from the AIA at 171 and 304 \AA\ pass-bands, were used and processed to level 1.5 using {\it aia\_prep.pro} to correct for solar north, plate-scale and alignment adjustment.
{Level 1 SWAP images were prepared using {\it p2sw\_prep.pro} with corrections for dark, flat-field, point spread function (PSF) deconvolution, despiking and the images were corrected.} SWAP observes the coronal emission at around the 174 \AA\ pass-band, with a larger FOV than AIA and at a cadence of $\approx$ 2 minutes. STEREO/EUVI images in the 304 \AA\ pass-band were used due to the instruments large FOV, and the relatively high cadence of 10 minutes that EUVI observes in this pass-band. {  Due to relatively lower cadence of 2 hours for 171 \AA\ pass-band of EUVI, those images were not used for our analysis unlike AIA.} The data was processed to level 1 using the {\it secchi\_prep.pro} routine which corrects for flat-field and bias, calibrates to physical units and normalizes the filter response to clear filter.}
AIA and EUVI images were rebinned to 1024$\times$1024 pixels for the generalization of the algorithm and for saving the processing time. 

{ We used the aforementioned pass-bands to observe several periods in the near maximum phase of solar cycle 24 when off-limb solar eruptions could be clearly identified visually in the 171 \AA\ pass-band of AIA on 2012-04-08, 2012-06-27 and 2012-08-31, in the AIA 304 \AA\ channel the data used was observed on 2012-04-08 and 2014-07-08, in EUVI-A and EUVI-B at 304 \AA\ pass-bands on 2013-05-13 and 2012-08-31 respectively and in SWAP 174 \AA\ pass-band on 2011-12-24, 2012-04-16, 2013-05-01, 2013-06-21 and 2014-08-24.}

\begin{figure}[!ht]
    \centering
    \begin{tikzpicture}[node distance=1.35cm] 
\node (start) [flowbox] {Radial filtering and disk masking};
\node (pro1) [flowbox, below of=start] {Cartesian to polar transformation of image};
\node (pro2) [flowbox, below of=pro1] {Fourier motion filtering to separate moving features};
\node (pro3) [flowbox, below of=pro2] {Dimensionality reduction of the polar maps};
\node (pro4) [flowbox, below of=pro3] {Generation of CME map};
\node (pro5) [flowbox, below of=pro4] {Intensity thresholding and labelling the regions in the CME map};
\node (pro6) [flowbox, below of=pro5] {Generation of height-time plot};
\node (pro7) [flowbox, below of=pro6] {Parabolic Hough transform to identify accelerating solar eruptions from height-time plot};
\node (pro8) [flowbox, below of=pro7] {Record solar eruptions properties};

\draw [arrow] (start) -- (pro1);
\draw [arrow] (pro1) -- (pro2);
\draw [arrow] (pro2) -- (pro3);
\draw [arrow] (pro3) -- (pro4);
\draw [arrow] (pro4) -- (pro5);
\draw [arrow] (pro5) -- (pro6);
\draw [arrow] (pro6) -- (pro7);
\draw [arrow] (pro7) -- (pro8);

\draw [decorate,decoration={brace,amplitude=5pt,mirror,raise=6pt},yshift=-2.6cm]
(3.5,0.65) -- (3.5,3.15) node [black,midway,xshift=2.1cm] {\footnotesize {\it
Intensity Enhancement}};
\draw [decorate,decoration={brace,amplitude=5pt,mirror,raise=6pt},yshift=-3.95cm]
(3.5,0.65) -- (3.5,1.8) node [black,midway,xshift=2.1cm] {\footnotesize
{\it Fourier Motion Filtering}};
\draw [decorate,decoration={brace,amplitude=5pt,mirror,raise=6pt},yshift=-7.95cm]
(3.5,0.65) -- (3.5,4.4) node [black,midway,xshift=3.35cm] {\footnotesize {\it
Automatic Identification of Solar Eruptions}};
\draw [decorate,decoration={brace,amplitude=5pt,mirror,raise=6pt},yshift=-12.0cm]
(3.5,0.65) -- (3.5,4.5) node [black,midway,xshift=3.35cm] {\footnotesize {\it
Application of Parabolic Hough Transform}};

\end{tikzpicture}
    \caption{Flowchart of the algorithm to automatically detect CMEs using the parabolic Hough transform.}
    \label{flowchart}
\end{figure}
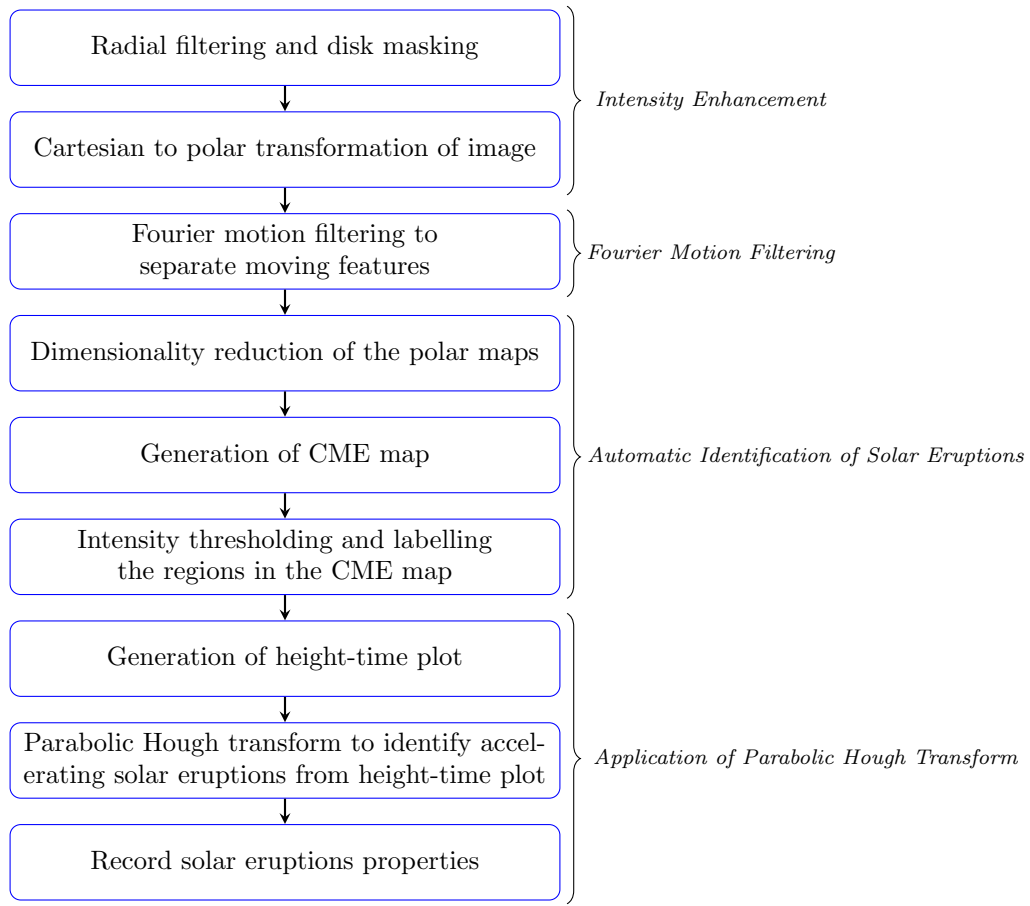

The SOHO/LASCO, STEREO/COR-2, and STEREO/HI-1 instruments image the outer corona where solar eruptions attain relatively constant velocities or have relatively moderate acceleration/deceleration profiles dictated by their interaction with the ambient solar wind, and are therefore ideally suited for detection by algorithms such as {  CACTus}, which uses a linear Hough transform \citep{NATG2000, Temmer2011}. 
As discussed in the introduction, the early evolution of solar eruptions show gradual and impulsive accelerations that are not seen when they reach the outer corona \citep{Bein2011}. {Solar eruptions in outer corona appear as straight lines in height-time (r-t) plots that are detected by CACTus using linear Hough transform.} On the other hand, solar eruptions propagating in inner corona appear as parabolic or higher order ($>2$) polynomial ridges in r-t plots. 
 Assuming that solar eruptions accelerate uniformly in the inner corona, we used parabolic Hough transform aided with Fourier motion filtering to detect and track solar eruptions in solar EUV images. This assumption is meant for simplicity in the automated detection. Automated detection of the 2nd order polynomial (parabola) is a first step towards the detection of higher order polynomials ($> 2$) which will be done in future studies. 
 The steps involved in CIISCO for detecting solar eruptions is outlined in the flowchart of Figure \ref{flowchart}.\\


 The method of detection employed by CIISCO is illustrated using SWAP Level 1 images taken on 2011-12-24, and the steps explained in the next few subsections.

\begin{figure}[!ht]   
   \centerline{\hspace*{0.05\textwidth}
               \includegraphics[width=0.7\textwidth,clip=]{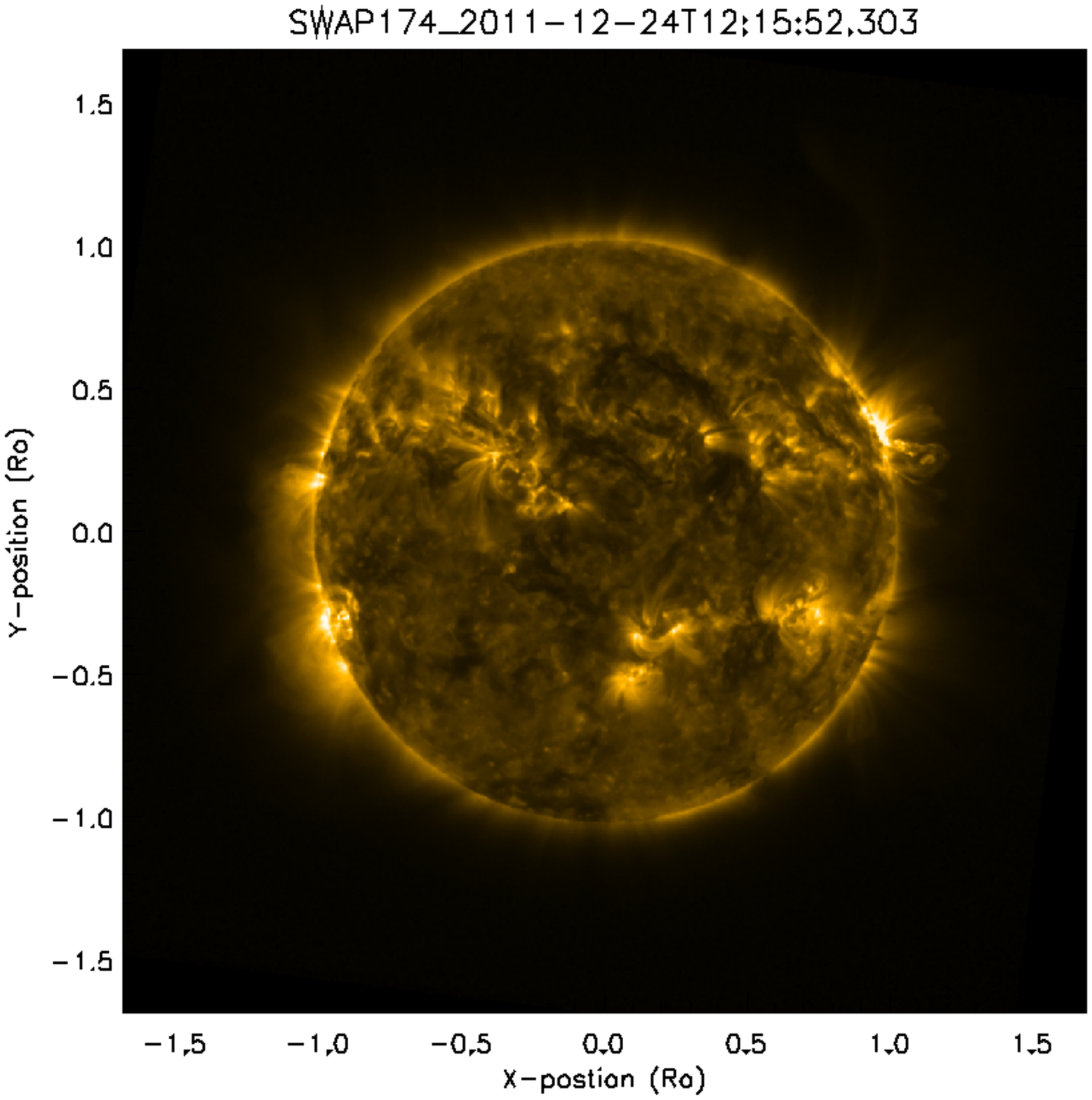}
               \hspace*{-0.075\textwidth}
               \includegraphics[width=0.7\textwidth,clip=]{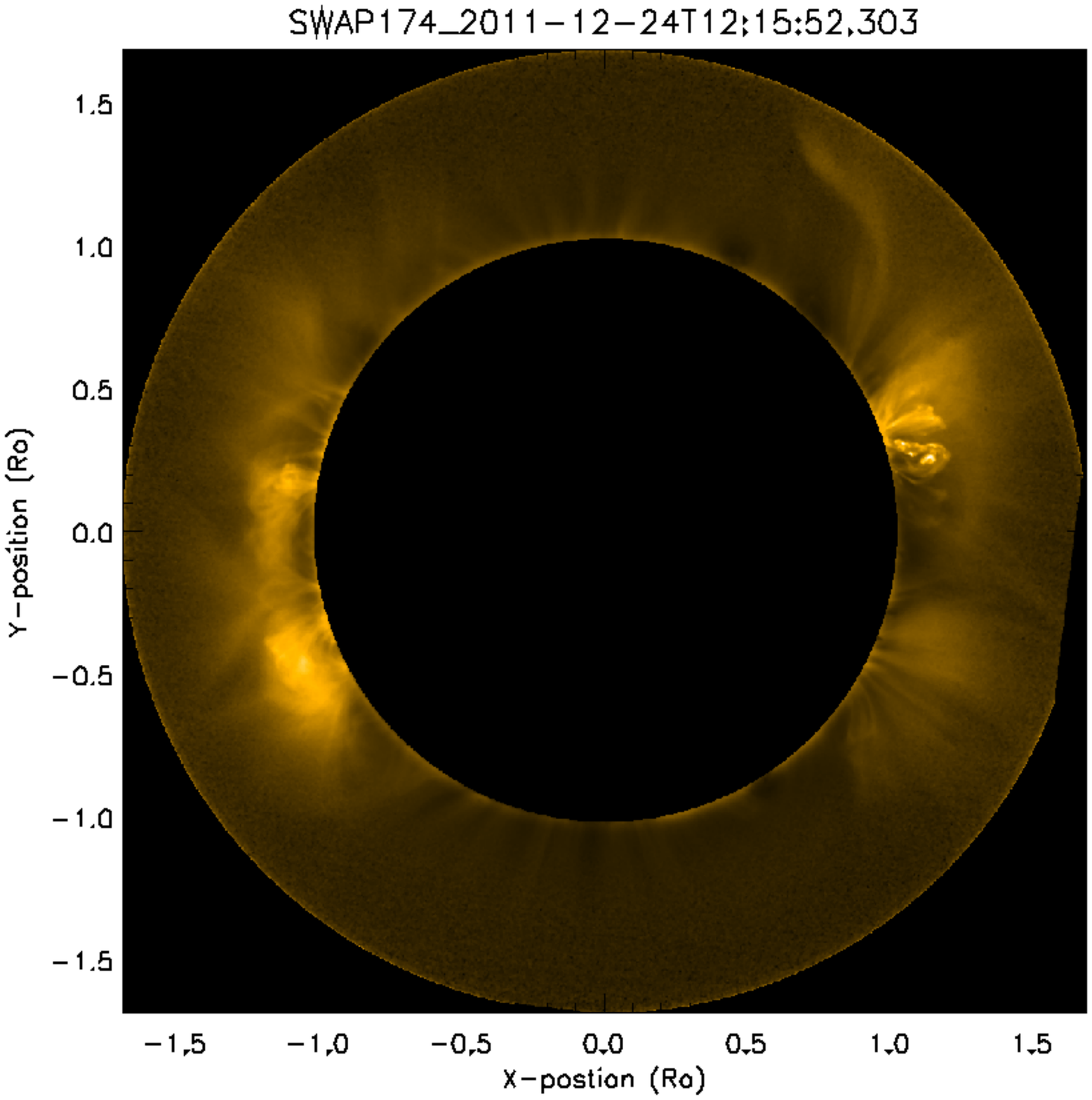}
              }
      \vspace{-0.01\textwidth}  
     \centerline{    
      \hspace{0.2\textwidth}  \color{black}{(a)}
      \hspace{0.58\textwidth}  \color{black}{(b)}
         \hfill}
     \vspace{0.005\textwidth}     
          
 \centerline{\hspace*{0.05\textwidth}
               \includegraphics[width=0.58\textwidth,clip=]{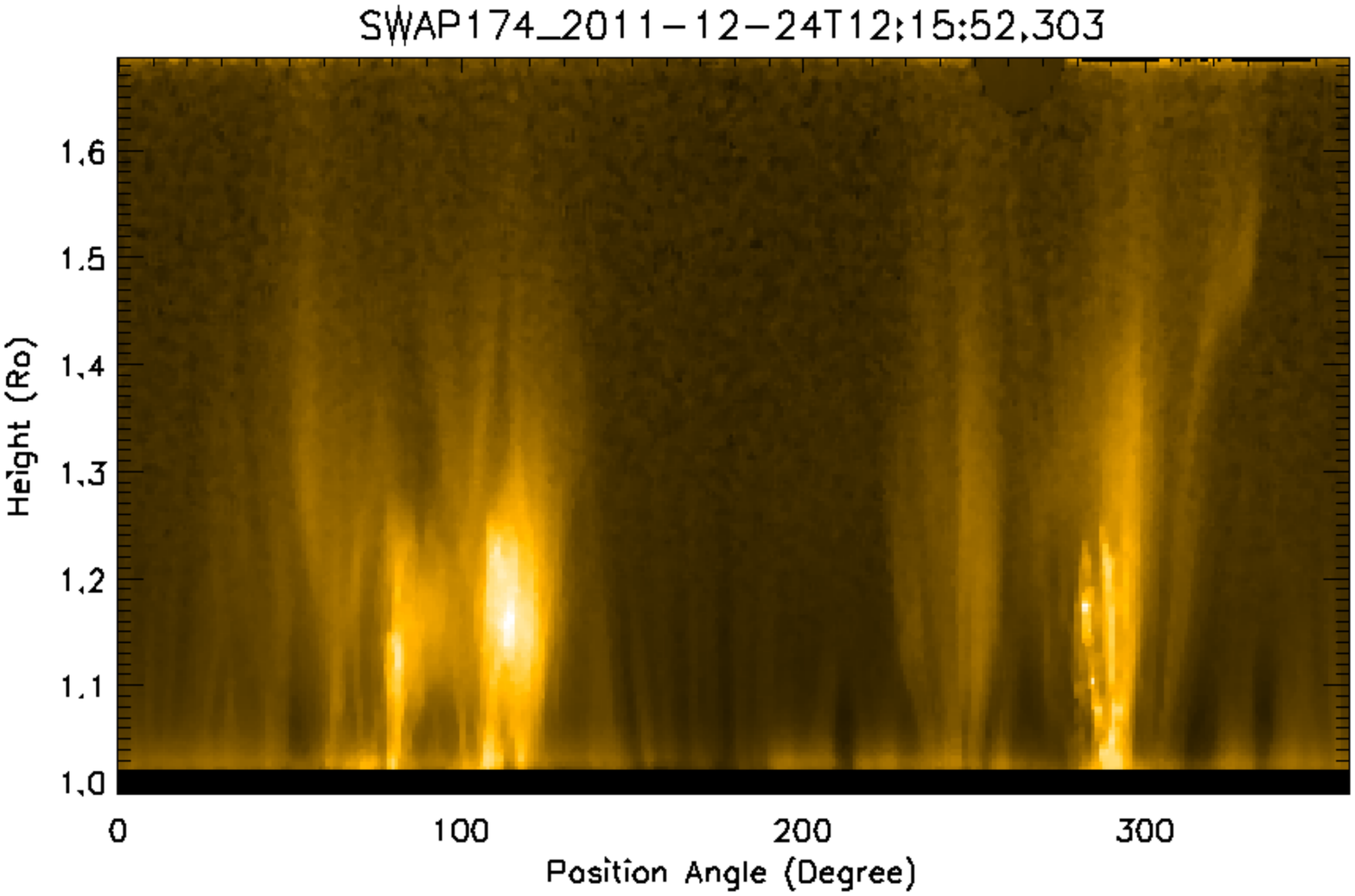}
               \hspace*{0.002\textwidth}
               \includegraphics[width=0.58\textwidth,clip=]{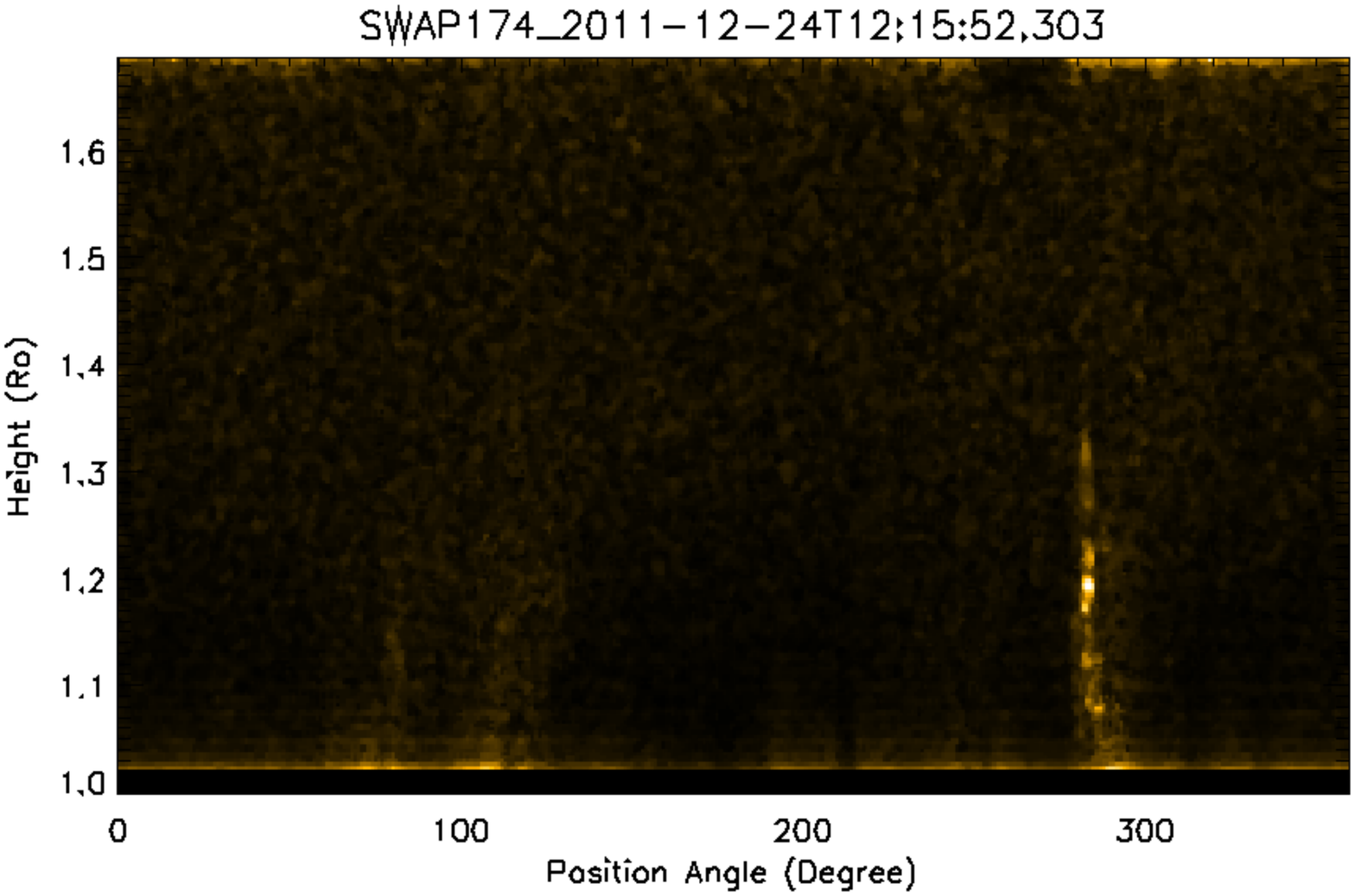}
              }
     \vspace{-0.01\textwidth}  
     \centerline{   
      \hspace{0.2\textwidth}  \color{black}{(c)}
      \hspace{0.58\textwidth}  \color{black}{(d)}
         \hfill}
     \vspace{0.01\textwidth} 
          
\caption{Preparation of a SWAP image to be used by CIISCO;(a) Level 1 SWAP image, (b) After radial filter and disk masking, (c) Polar transformation of (b), (d) After Fourier motion filtering. CME signature can be seen at PA $\approx$290$^\circ$}.
   \label{fig:algofig1}
\end{figure}

\subsection{Intensity Enhancement}
\label{prep}
 
 As density, and consequently intensity, in the corona decreases radially with distance from the solar limb, coronal features appear fainter at larger distances. In order to enhance the coronal intensity in off-limb corona, a radial filter was applied \citep{Morgan06}. 
First we create a background image taking an average of the lowest 3\% intensities at each pixel of all images of the dataset, following  \cite{DeForest14}. A radial intensity profile of this image was generated taking a radial cut at the polar region with fixed width of three pixels and then averaging their intensities. To avoid the errors created by bad pixel intensities, this array of radial intensity profile was further smoothed out by 10 pixels in the radial direction.
The polar regions were chosen as they have fewer foreground structures (e.g. loops).  This radial profile was used to make an azimuthally uniform background that is used as a radial filter. {Each of the images in dataset were then divided by this radial filter.} By applying these techniques, coronal structures were clearly seen to greater distances (see Figure \ref{fig:algofig1}b), especially when compared to Figure \ref{fig:algofig1}(a) which is unfiltered. As we were interested in off-disk features, we blocked the solar disk by applying an artificial mask up to 1.02 R$_{\odot}$ 
The images were then converted to polar ($\theta$-r) coordinate system, where $\theta$ is the position angle (PA) in counter-clockwise direction measured from solar north, and $r$ is the distance from center of the Sun projected in plane of sky. 
 Figure \ref{fig:algofig1}(c) shows the polar transformed image.

{  \subsection{Fourier Motion Filtering}  \label{fomo}
The inner corona shows many static and quasi-static structures such as coronal loops, prominences, etc.  near the solar limb. Any changes in their intensity or morphology may appear as bright ridges in time difference images which may create false detections of solar eruptions. To avoid this, we used the Fourier motion filtering technique of \citep{DeForest14} to separate solar eruptions from these background structures. We generated height-time (r-t) plots at each PA and took the Fourier transform giving spatial frequency (k) along x-axis and temporal frequency ($\omega$) along y-axis. In k-$\omega$ plot the first and third quadrants indicate the inbound features whereas the second and fourth correspond to outbound features.
Any outward moving structure will have a positive gradient in r-t plot and would correspond to velocities given by slopes of lines falling in second and fourth quadrants of k-$\omega$ plot. By masking the first and third quadrants of the k-$\omega$ plot, the inbound and outbound solar eruptions were separated. The static features which do not change over time and space were removed by masking out the low frequency components in k-$\omega$ space. {  It was found that for the given resolution in Fourier space, the cut-off speed corresponding to single pixel below which detection will not happen in AIA, SWAP and EUVI are 14 km s$^{-1}$, 19 km s$^{-1}$ and 76.45 km s$^{-1}$ respectively. It has been found that during the slow-rise phase, the CMEs have speed in the range of 5 - 80 km s$^{-1}$ \citep{Zhang_2001}, most of which falls above the cut-off speed limit in the Fourier space for AIA and SWAP observations. The lower cadence of EUVI images limits to capture such slow rise-phase of CME propagation.} The inverse Fourier transform of the masked k-$\omega$ data gives the Fourier filtered r-t plot which were stacked at each PA to generate polar images with only outward moving features present. Figure \ref{fig:algofig1}(d) shows the coronal structures after Fourier motion filtering has been applied and the data filtered for outward moving structures. }

\begin{figure}[!ht]   
   \centerline{\hspace*{0.05\textwidth}
               \includegraphics[width=0.58\textwidth,clip=]{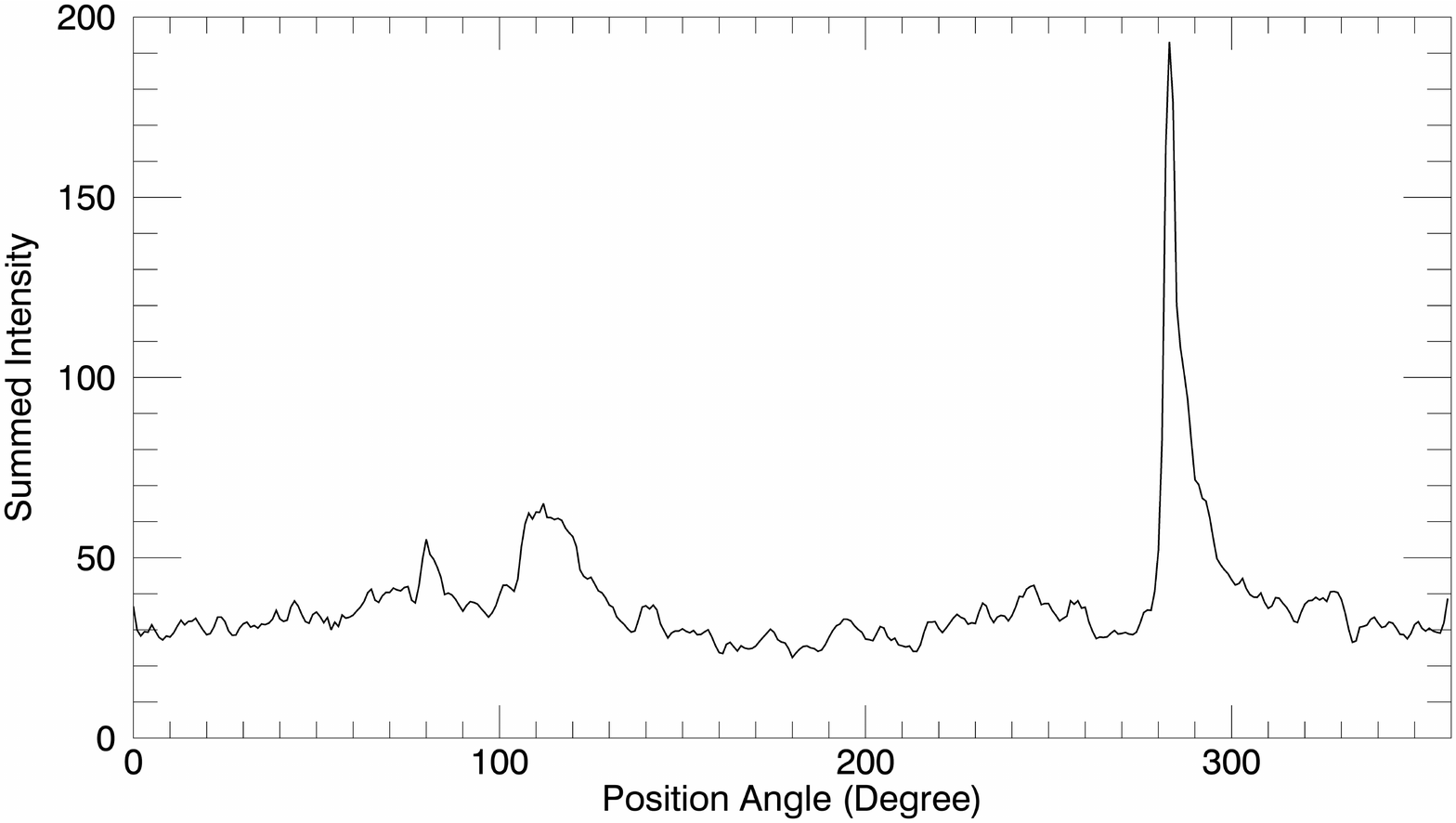}
               \hspace*{0.002\textwidth}
               \includegraphics[width=0.58\textwidth,clip=]{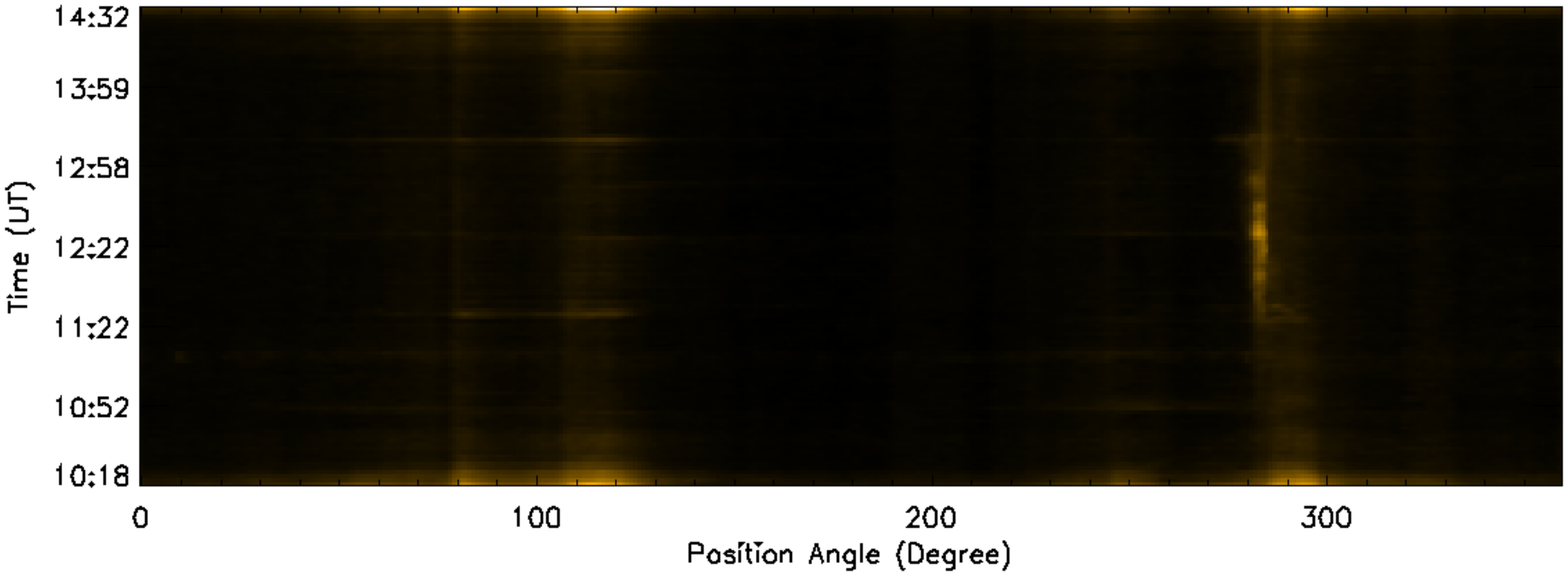}
              }
      \vspace{-0.01\textwidth}  
     \centerline{    
      \hspace{0.2\textwidth}  \color{black}{(a)}
      \hspace{0.58\textwidth}  \color{black}{(b)}
         \hfill}
     \vspace{0.005\textwidth}     
          
 \centerline{\hspace*{0.05\textwidth}
               \includegraphics[width=0.58\textwidth,clip=]{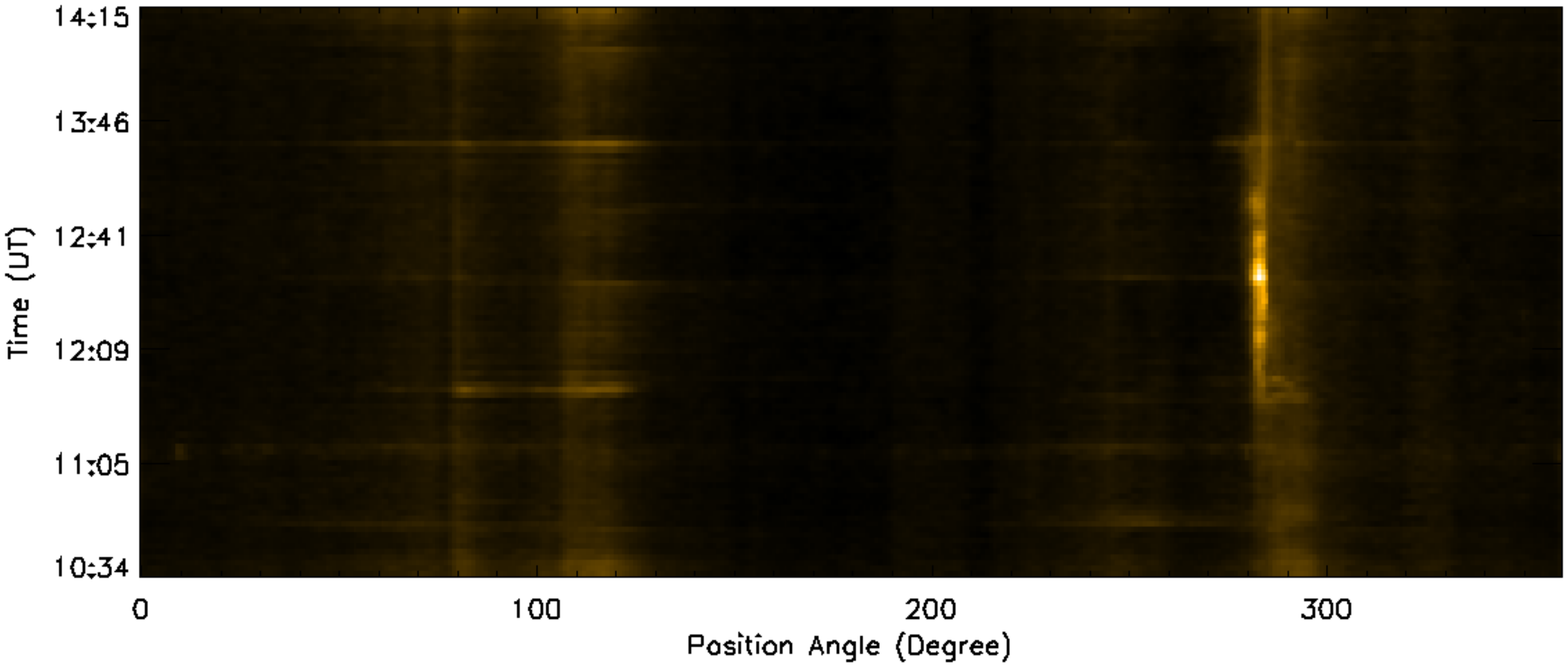}
               \hspace*{0.002\textwidth}
               \includegraphics[width=0.58\textwidth,clip=]{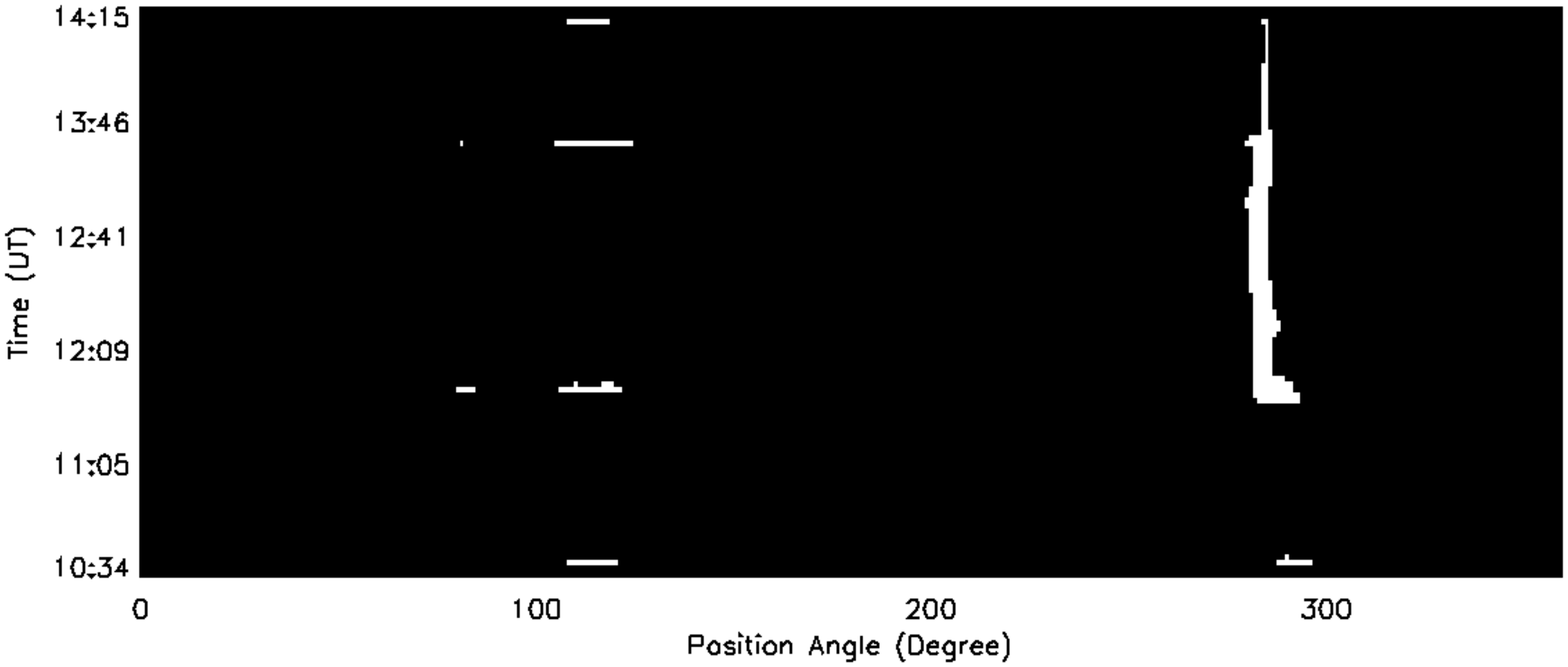}
              }
     \vspace{-0.01\textwidth}  
     \centerline{   
      \hspace{0.2\textwidth}  \color{black}{(c)}
      \hspace{0.58\textwidth}  \color{black}{(d)}
         \hfill}
     \vspace{0.01\textwidth} 

\centerline{\hspace*{0.05\textwidth}
               \includegraphics[width=0.58\textwidth,clip=]{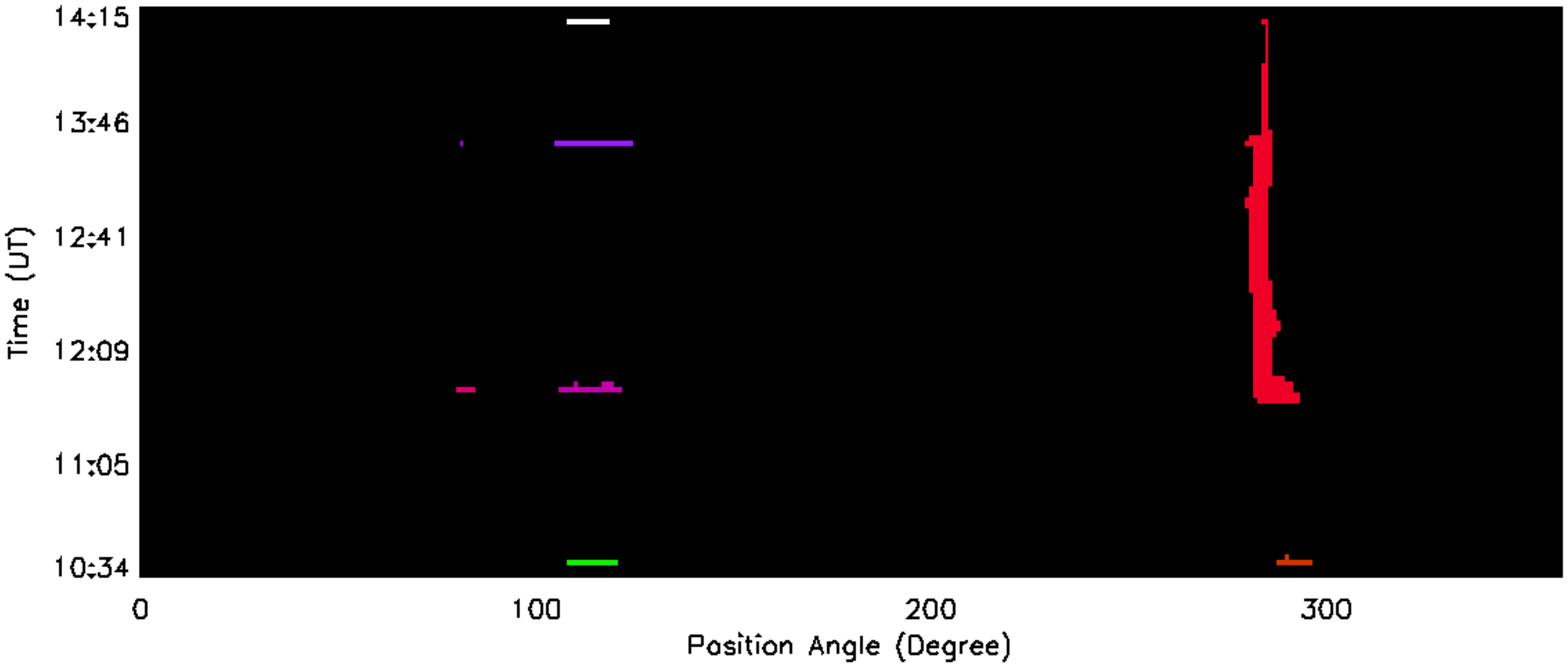}
               \hspace*{0.002\textwidth}
              }
     \vspace{-0.01\textwidth}  
     \centerline{   
      \hspace{0.5\textwidth}  \color{black}{(e)}
         \hfill}
     \vspace{0.01\textwidth} 
          
\caption{Outlining the automatic identification of solar eruptions in the Fourier motion filtered images; (a) 1D intensity plot created by summing up intensities along each PA, (b) CME map created after stacking 1D intensity plots in time, (c) Cropped CME map, (d) CME map after intensity thresholding, (e) After labelling regions with different colors.}
   \label{fig:algofig2}
\end{figure}

\subsection{Automatic Identification of Solar Eruptions} \label{cmemap}
A quick inspection of Figure \ref{fig:algofig1}(d) reveals a {solar eruption}, seen as a bright structure, spanning some PAs near 290$^\circ$. To identify the solar eruptions, the Fourier filtered polar images were converted to 1D arrays by summing up intensities along radial directions at each PA as shown in Figure \ref{fig:algofig2}(a), (similar to SEEDS). Since solar eruptions appear brighter than the background, it corresponds to a higher intensity in 1D intensity plots. 
Each polar image was integrated along the radial direction to create 1D arrays and stacked with-respect-to time, to generate so called CME map with time along vertical axis (Figure \ref{fig:algofig2}(b)). 
However, artifacts can be seen at the top and bottom of the CME map (approximately the first and last 10 images). These are believed to be created by the Fourier filtering technique, which also introduced faint ringing patterns in the CME map. These artifacts are bright enough to trigger false detection and are therefore removed from the map (see Figure \ref{fig:algofig2}(c)).
The cropped CME map is then passed through an intensity threshold algorithm using the relation,
\begin{equation} \label{thresh}
    \mathrm{Imap_{th}} = \mathrm{mean(cmemap)}+f \times \mathrm{stddev(cmemap)}, 
\end{equation}
where $Imap_{th}$ is the binary CME map after applying an intensity threshold to the cropped CME map ({\it cmemap}), f is an adjustable constant parameter dependent on the instrument response, which may vary with instrument and pass-band. 
{ The value of f was determined to be 3.25 by running CIISCO on various different eruptions in order to capture the most eruptions with minimal false detections whilst maximising the detection efficiency by comparing with manual detection.}
Figure \ref{fig:algofig2}(d) shows the CME map after intensity thresholding. Different regions obtained in $Imap_{th}$ are then labelled and marked with different colors, as shown in Figure \ref{fig:algofig2}(e). Regions with widths greater than 5$^\circ$, and persisting for at least three frames were considered to be the signatures of possible solar eruptions. {These thresholds applied on the CME map regions reduced the possibility of false detection in subsequent steps.} An estimate about the angular extent of the possible solar eruptions were made by determining the minimum and maximum PAs of these regions from the CME map. {It can be seen from Figure \ref{fig:algofig2}(e) that the region labelled in red at PA $\approx$290$^\circ$ satisfies both these criteria and can is therefore a potential {solar eruption} detection.
}

\subsection{Application of Parabolic Hough transform}   \label{pht}
\begin{figure}[!ht]   
   \centerline{\hspace*{0.05\textwidth}
               \includegraphics[width=0.4\textwidth,clip=]{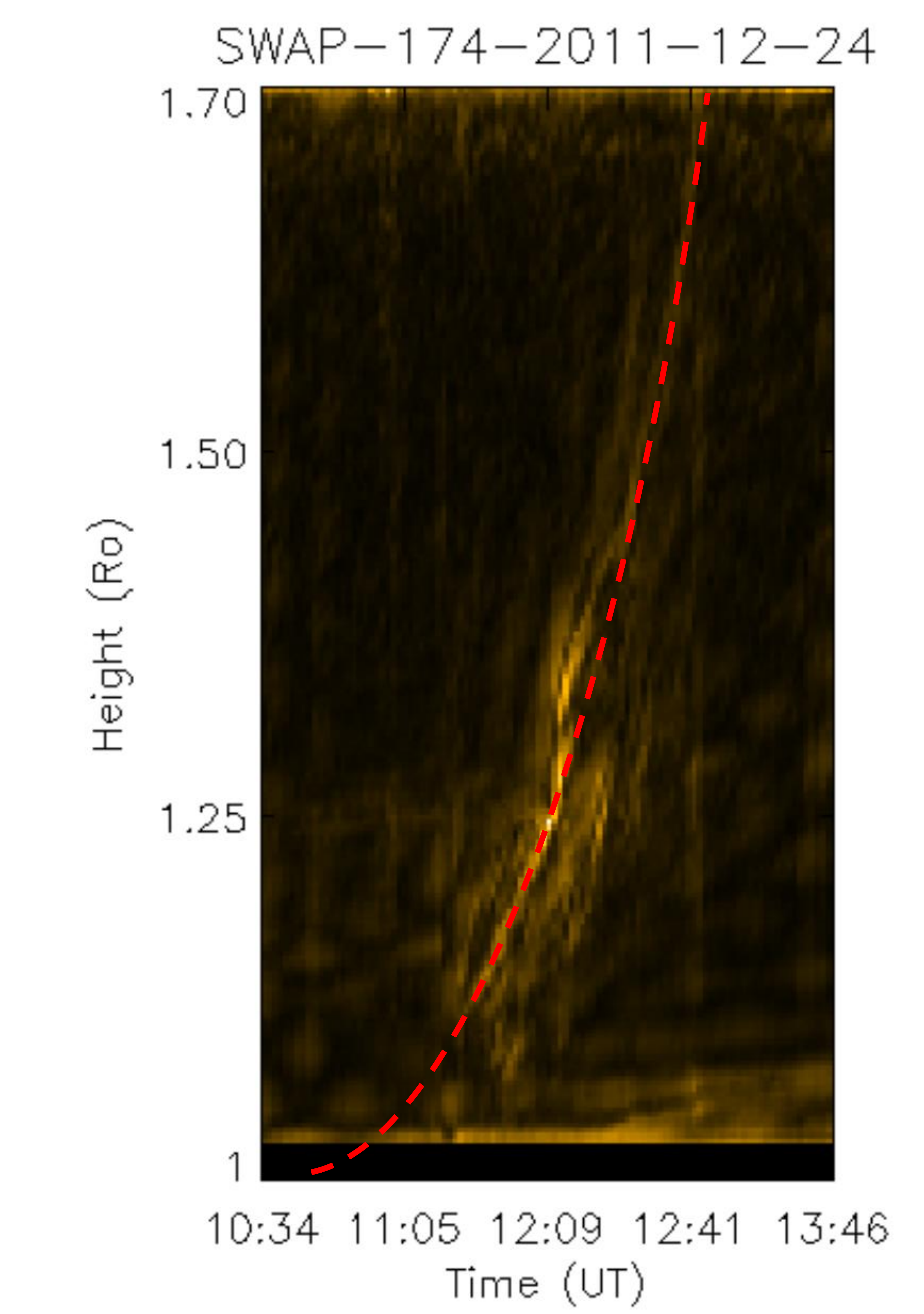}
               \hspace*{0.002\textwidth}
               \includegraphics[width=0.6\textwidth,clip=]{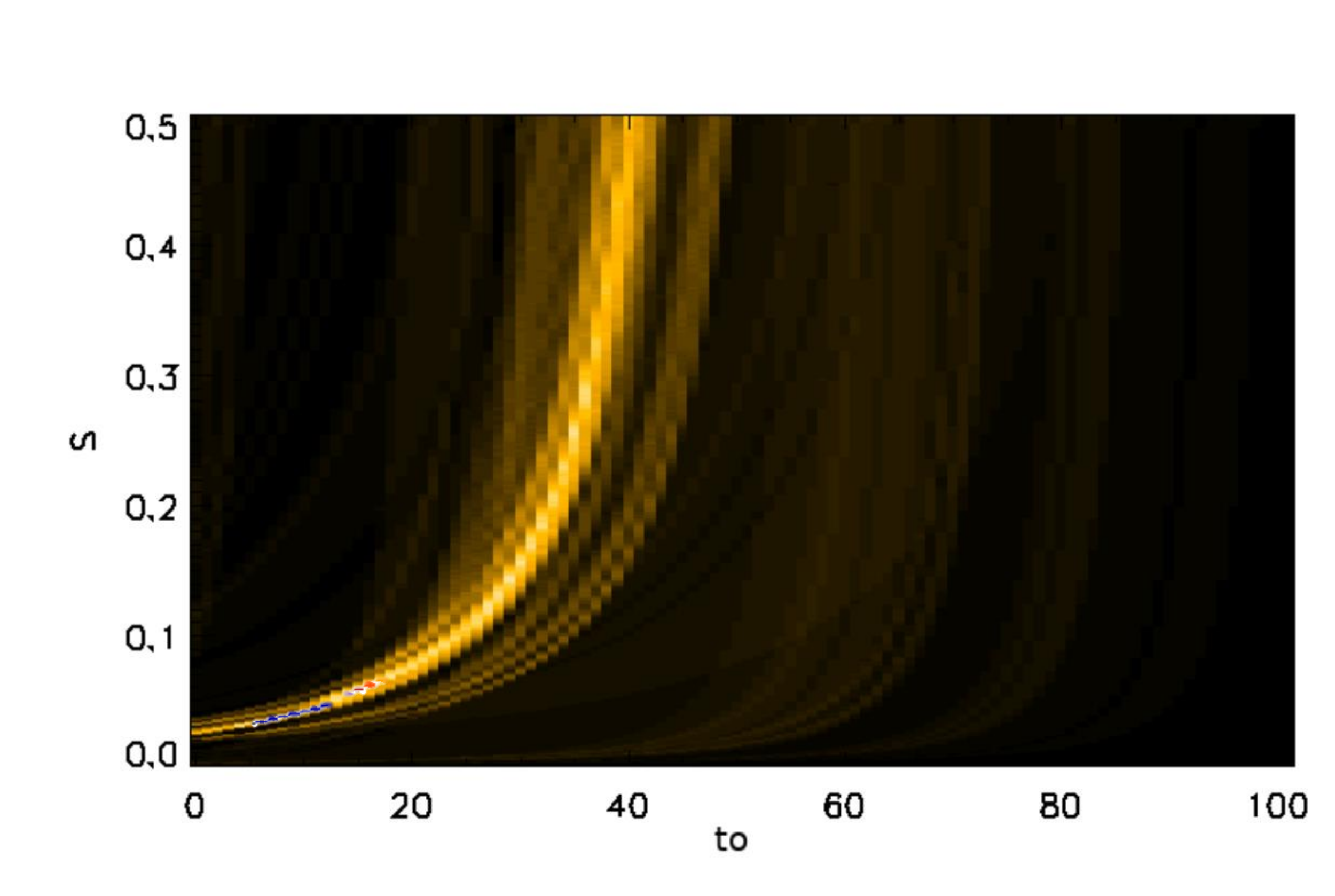}
              }
          

\caption{Application of parabolic Hough transform, (a) Height-time plot of {  solar eruption} location identified from Figure \ref{fig:algofig2}(e) with identified parabola over-plotted in dashed line, (b) Accumulator matrix created using $t_{\circ}$ and $S$ parameters with intensity threshold and labelled region shown with blue and red colors over-plotted.}
   \label{fig:algofig4}
\end{figure}

Once a probable {solar eruption} is identified, the next step is to track it in both temporal and spatial domain. {In a recent work, the centroid of prominences are used to track them in successive frames rather than leading edge of the eruption, for the SDO/AIA 304 \AA\ PE catalog mentioned in Section \ref{sec:intro} \citep{Yashiro2020arXiv}.} The existing  CME auto-detection methods \citep{Robbrecht04, Olmedo08, Pant2016, Byrne12} identifies the leading edge of the CMEs to derive their characteristics. However, the leading edge of the EUV eruptions gets distorted on their propagation outwards making it difficult to identify them satisfactorily in subsequent frames. Therefore, we tracked solar eruptions as a whole in the EUV images. For this we found the difference of maximum and minimum PAs from CME map of the identified solar eruptions, that gives an idea of the maximum width of the observed erupting structure. We then summed up the intensities along PAs of each motion filtered polar image within this width at each height.
Summing up the intensities also enhances the signal to noise ratio (SNR). Such arrays of subsequent images are then stacked in time to create r-t plots for the identified solar eruptions. The outward moving feature appears as a bright ridge in the r-t plot (see Figure \ref{fig:algofig4}(a)) and the identification of this ridge provides a representation of the tracked eruption.

Previous studies and observations have shown that most of the solar eruptions show acceleration in the inner corona \citep{Bein2011}. Thus {due to acceleration} they appear as parabolic ridges in r-t plots that can also be seen from Figure \ref{fig:algofig4}(a). For a general form of parabola, a hidden parameter $\theta$ determines its orientation with-respect-to horizontal axis. As solar eruptions start with zero velocity, the slope of parabola (velocity) at r$_{\circ}$ has to be zero. The slope of the parabola (velocity) at the origin (r$_{\circ}$, t$_{\circ}$) is zero if $\theta$=90$^\circ$. The parabolic ridges in r-t plots will then have form,
\begin{equation} \label{parabola}
    r-r_{\circ} = S(t-t_{\circ})^2 ,
\end{equation}
where r$_{\circ}$ and t$_{\circ}$ are the spatial and temporal starting points of the parabolic ridges in the r-t plots, {  $S$ is a coefficient defining the curvature of the parabola, and hence representing the acceleration of the {  solar eruption} and finally $\theta$ defines the orientation angle of the parabola.} 

\begin{figure}
    \centering
    \includegraphics[width=0.8\textwidth,clip=]{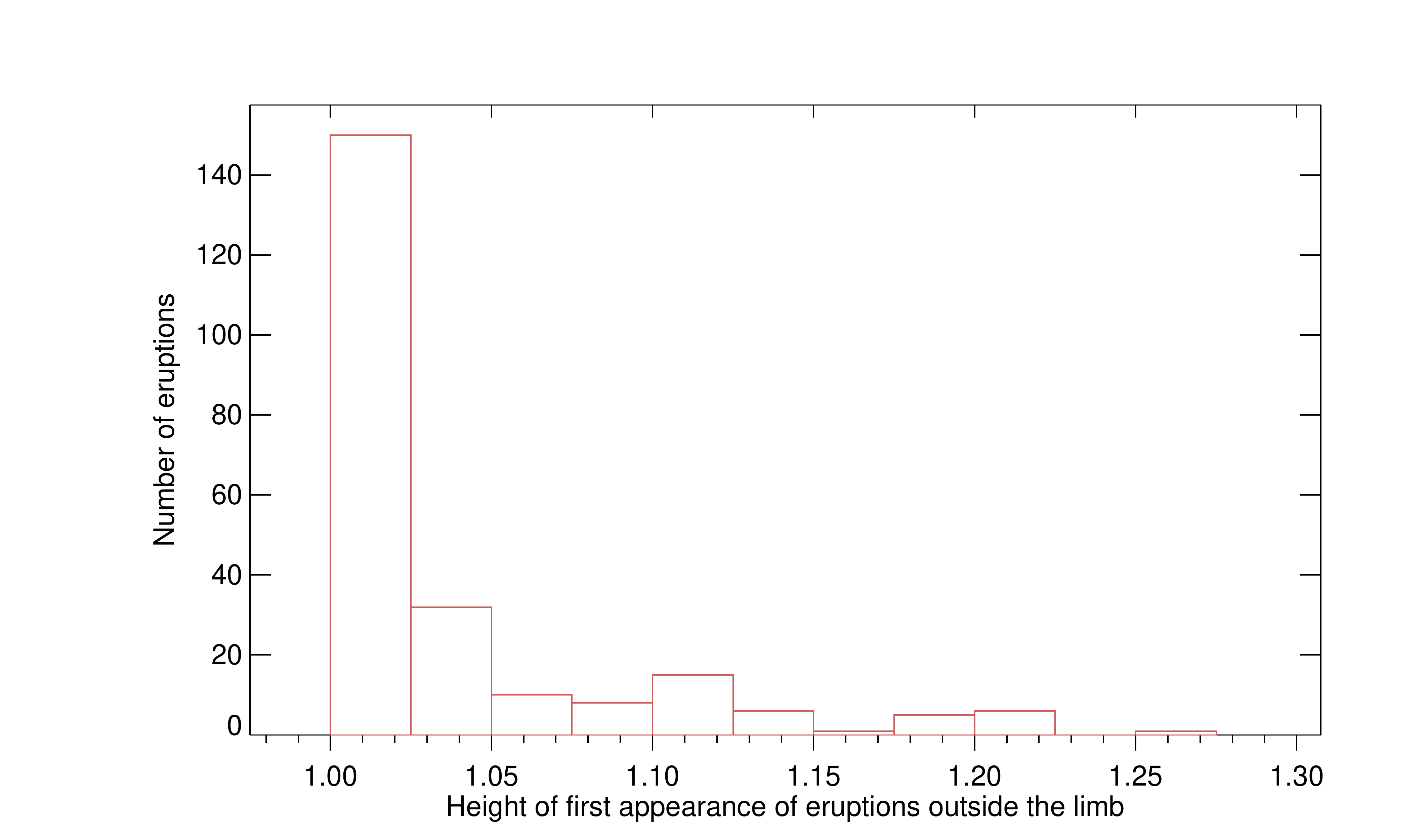}
    \caption{Distribution of heights of first appearance of eruptions outside the limb observed by SDO/AIA in 171 and 304 \AA\ passbands from 2012-01-01 to 2012-04-30.}
    \label{fig:hist_erupt}
\end{figure}

The detection of a {solar eruption} represented by the parabolic ridge in images using the Hough transform requires all four parameters, t$_{\circ}$, r$_{\circ}$, {\it S} and $\theta$ \citep{Ballard81}. Identifying the position of these parameters in 4D parameter space to define a parabola is computationally expensive, but can be improved by reducing the degrees of freedom in Equation \ref{parabola}. {As we aim to detect the off-disk eruptions, we take the value of r$_{\circ}$ where the eruption first appear outside the solar disk. To determine the value of this parameter we measured the height of first appearance of limb and near limb eruptions observed in SDO/AIA images taken at 171 and 304 \AA\ from 2012-01-01 to 2012-04-30 using Jhelioviewer \citep{JHelio}. We identified 234 eruptions during this period and plotted a distribution as shown in Figure \ref{fig:hist_erupt}. Out of these 234 eruptions, $\approx$ 64\% show a starting height in the bin of 1.00 to 1.025 R$_{\odot}$. Therefore, we fixed r$_{\circ}$ to the lower limit of this bin as 1 R$_{\odot}$.} This assumption holds good for the eruptions occurring near the solar limb rather than for those occurring near the Sun centre. Thus, we reduce the 4D ({\it S}, t$_{\circ}$, r$_{\circ}$,  $\theta$) problem to that of a 2D ({\it S}, t$_{\circ}$, r$_{\circ}$=1 R$_{\odot}$, $\theta$=90$^\circ$) problem.  Equation \ref{parabola} can be further written in the form,

\begin{equation} \label{eqacc}
    t_{\circ} = t-\sqrt{\frac{1}{S}(r-r_{\circ})}.
\end{equation}

To identify parabolic ridges in Figure \ref{fig:algofig4}(a) defined by Equation \ref{parabola}, a 2D array called the accumulator matrix is made using Equation \ref{eqacc} with horizontal and vertical axes as $t_{\circ}$ and $S$ respectively.
From the r-t plot, for each value of t, $t_{\circ}$ is calculated iteratively by varying the values of S. The value at ($t_{\circ}$, S) in accumulator matrix is increased by one whenever a pixel corresponding to  {  parabolic bright ridge} is detected in r-t map. The resulting accumulator matrix would appear as shown in Figure \ref{fig:algofig4}(b). {  It can be seen that the accumulator matrix consists of parabolas with different intensities. Each point of accumulator matrix gives a pair of $t_{\circ}$ and S values corresponding to different parabolic trajectories in the r-t plot. However, the one corresponding to the brightest parabolic ridge of Figure \ref{fig:algofig4}(a) will have pixel values increased by most of the iterations. As a result that pixel in accumulator matrix will appear the brightest. This is the basic principle of the Hough transform to identify a feature from a noisy background that we have utilised to identify parabolas. Ideally, the pixel in the accumulator matrix with the maximum intensity should correspond to the parabolic ridge to be identified in the r-t plot. The coordinates of this pixel will give the values of $t_{\circ}$ and S defining the identified ridge. As it can be seen in Figure \ref{fig:algofig4}(a), the bright parabolic ridge corresponding to the solar eruption is a group of pixels with certain width and different intensities along the width, it will eventually lead to a group of pixels in the accumulator matrix which correspond to this ridge. The set of brightest pixel coordinates in the accumulator matrix shown in Figure \ref{fig:algofig4}(b) provides the $t_{\circ}$ and S parameters defining this parabolic ridge. Therefore, an intensity threshold is applied to the accumulator matrix with threshold of 90 $\%$ of the maximum intensity to identify the combination of $t_{\circ}$ and S for the desired identification. This is followed by morph closing operation to avoid values being missed after thresholding.} The accumulator matrix after intensity thresholding and the morph closing operation is shown with labelled regions in blue and red colors over-plotted on Figure \ref{fig:algofig4}(b).
 {  As the parabolic ridges have a width of a few pixels, we have taken the median of all values of ($t_{\circ}$, S) of the connected pixels of the thresholded accumulator matrix with a width greater than 10 pixels.} This may result in a family of parabolas being identified for a given {  solar eruption} in r-t plots if more than one regions satisfying this criterion is identified. Parabolas close in time are considered to be identified for the same {  solar eruption} in the r-t plot. 
Figure \ref{fig:algofig4}(a) also shows the identified parabola over-plotted in red dashed line on parabolic ridge. Note that this particular set of thresholds in Hough space is successful in identifying {  solar eruption} from other spurious parabolic signatures.

\subsection{Determination of {Solar Eruption} Properties} \label{cmeprop}
Once a {solar eruption} has been identified using the parabolic Hough transform, several characteristics, describing each {eruption}, are extracted. {These characteristics are derived by detecting and tracking the eruption as a whole and not just the leading front.} These include: 
\begin{enumerate}
    \item {\it Estimation of central position angle (CPA)}: The CPA of a {  solar eruption} is calculated from the CME map as the midpoint of the maximum width of the corresponding labeled region in the direction of PA.
    \item {\it Estimation of {  solar eruption} onset time (t$_{\circ}$)}: The onset time is estimated from the parameter t$_{\circ}$ in Equation \ref{eqacc}. As a set of parabolas are identified for a given {  solar eruption}, we take the mean of t$_{\circ}$ of all these parabolas to estimate the onset time of {  solar eruption}.
    \item {\it Estimation of {  solar eruption} kinematics}: The average {  apparent} velocity (v) was determined by calculating the slope of the line joining the two extreme points in the identified parabola. {  The average apparent} acceleration (acc) of the {  solar eruption} was determined from the constant $S$ of the parabola. Comparing the equation of motion,
\begin{equation} \label{motion}
    s = s_{\circ} + ut + \frac{1}{2}at^2 ,
\end{equation}
and Equation \ref{parabola}, we found that the apparent acceleration was 2$S$ with an assumption that initial velocity of {  solar eruption}, $u$, was zero. We determined the average {  apparent} velocity and acceleration for each of the parabolas in the family and take the mean value to be recorded for {  solar eruption} kinematics.
\end{enumerate}

For the example of SWAP images taken on 2011-12-24 we obtained that the {  solar eruption} onset time is  11:28UT, CPA is 287$^\circ$, average {  apparent} velocity is 102 km s$^{-1}$ and average {  apparent} acceleration is 67 m s$^{-2}$ after the application of CIISCO.

\section{Results} \label{sec:results}
\begin{figure}[!ht]   
   \centerline{\hspace*{0.05\textwidth}
               \hspace*{0.002\textwidth}
               \includegraphics[width=0.4\textwidth,clip=]{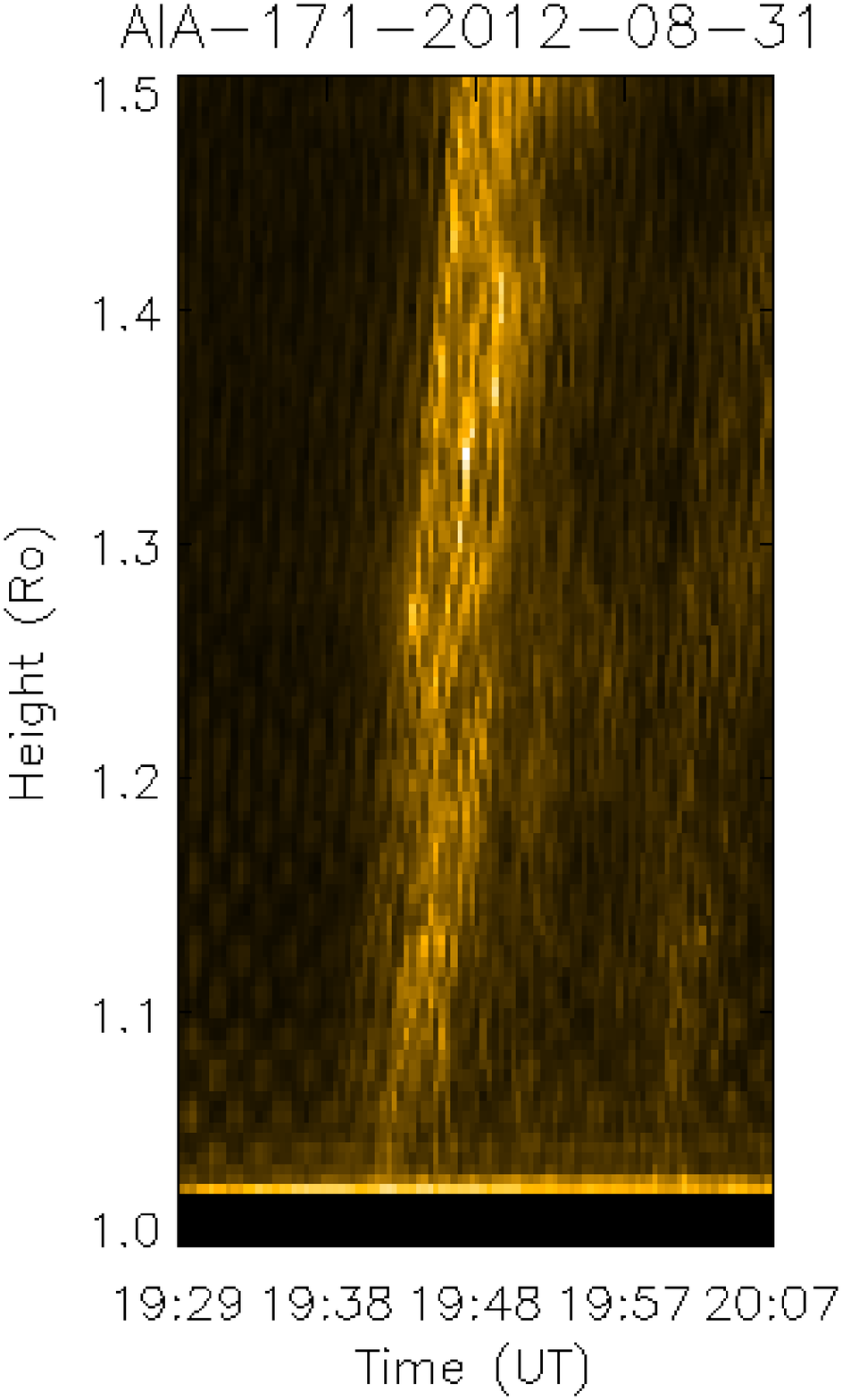}
               \hspace*{0.002\textwidth}
               \includegraphics[width=0.4\textwidth,clip=]{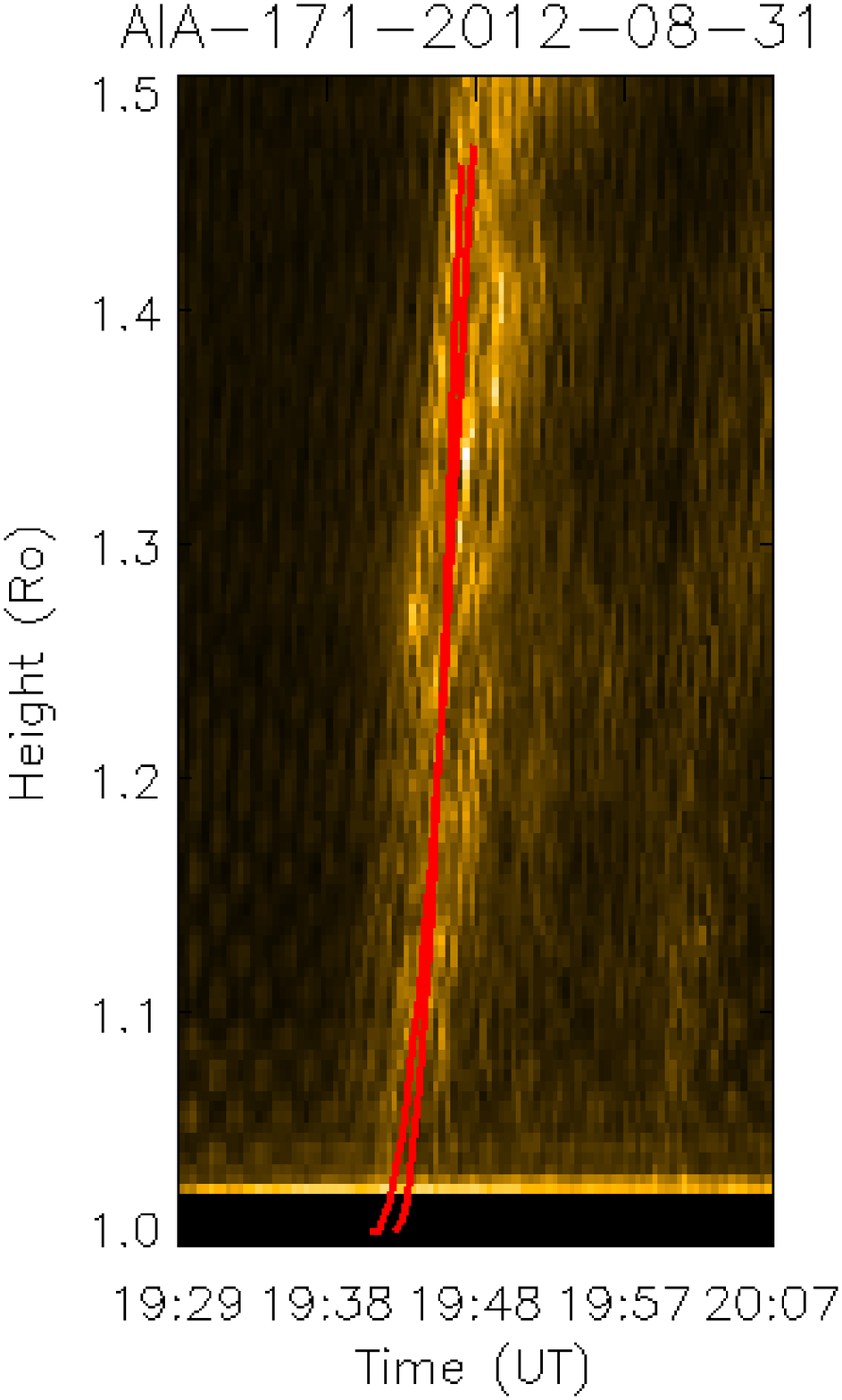}
              }
      \vspace{-0.01\textwidth}  
     \centerline{    
      \hspace{11em}  \color{black}{(a)}
      \hspace{13em}  \color{black}{(b)}
         \hfill}
     \vspace{0.005\textwidth}     
          
   \centerline{\hspace*{0.05\textwidth}
               \includegraphics[width=0.4\textwidth,clip=]{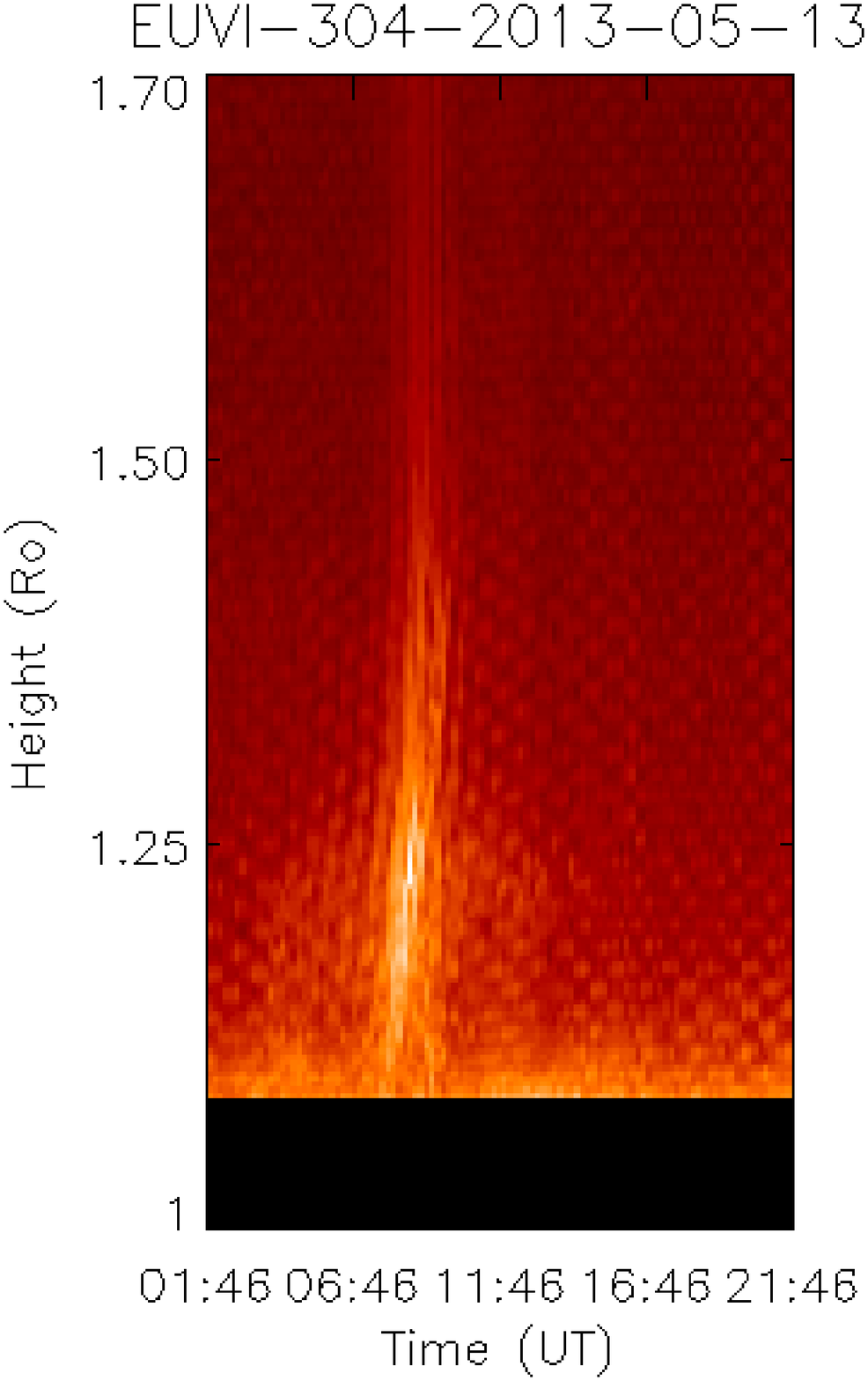}
               \hspace*{0.002\textwidth}
               \includegraphics[width=0.4\textwidth,clip=]{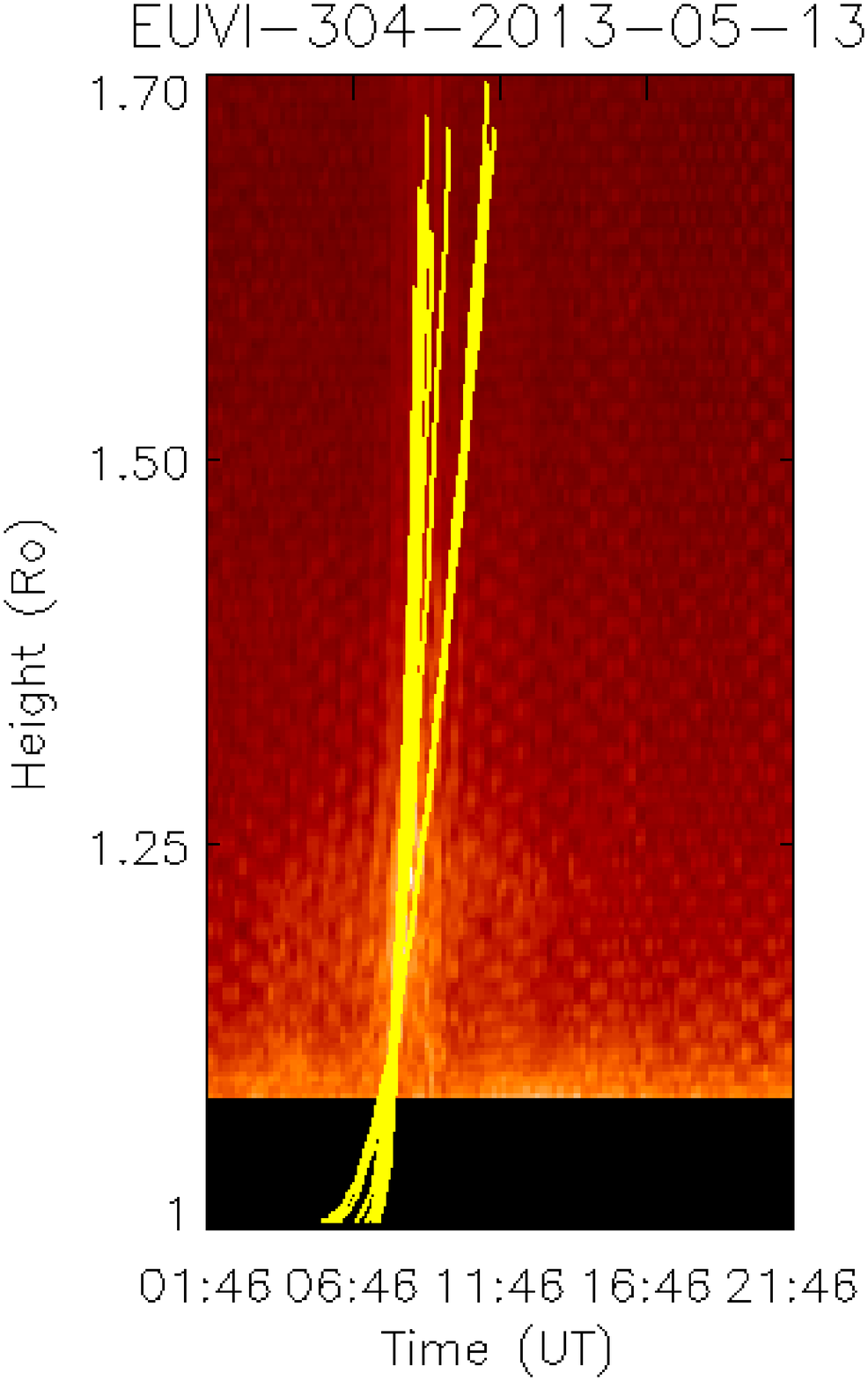}
               \hspace*{0.002\textwidth}
              }
      \vspace{-0.01\textwidth}  
     \centerline{    
      \hspace{11em}  \color{black}{(c)}
      \hspace{13em}  \color{black}{(d)}
         \hfill}
     \vspace{0.005\textwidth}     

\caption{r-t plots and identified {solar eruption} parabolas generated by applying CIISCO to EUV images. Panels (a) and (b) show the {  solar eruption} identified in AIA-171 \AA\ observations on 2012-08-31 without and with (respectively) identified parabolas over plotted. Similar to panels (a) and (b); (c) and (d) show observations from  EUVI-A-304 on 2013-05-13.}
   \label{fig:application}
\end{figure}

We applied CIISCO to EUV images of SWAP, AIA and EUVI over several short time periods. The results can be seen in Figure \ref{fig:application}, where solar eruptions are identified as parabolic ridges in r-t plots in EUV images. The results are summarized in Table \ref{table1}. 
The following properties of a {solar eruption} are determined by CIISCO: the solar eruption onset time (t$_{\circ}$), the central position angle (CPA), the average {  apparent} velocity (v) and the average {  apparent} acceleration (acc). The range of {  apparent} velocities and accelerations obtained from CIISCO are recorded and indicated by their extremes as minv, maxv, minac and maxac respectively. { For completeness, we have also manually identified the eruptions, tracked their leading front in consecutive images and calculated the average {apparent} speed (vm) and average {apparent} acceleration (acm) measured in the plane of sky for comparison with CIISCO. A contour was manually fitted to the outermost boundary of the eruption that could be visually identified. To take the fitting accuracy into account, the height of the leading edge was estimated by taking the average of all those points whose heights are greater than 0.95 times the maximum height obtained by fitting the contour.}
 {  To elucidate Table \ref{table1}, the used instrument, pass-band of observation, and date of solar eruption are included. A serial number of the eruption on a particular date is given in column 3. The last column indicates if the detection was a false positive, where no eruption was observed following manual inspection. False positives were often triggered by rising plasma motions in loops or the appearance of moving features on the limb created by solar rotation also contribute to false category. The remaining detections are treated as true if a radially outward moving feature is present in images and { has} been visually identified.}

\begin{figure}[!ht]   
   \centerline{\hspace*{0.05\textwidth}
               \includegraphics[width=0.6\textwidth,clip=]{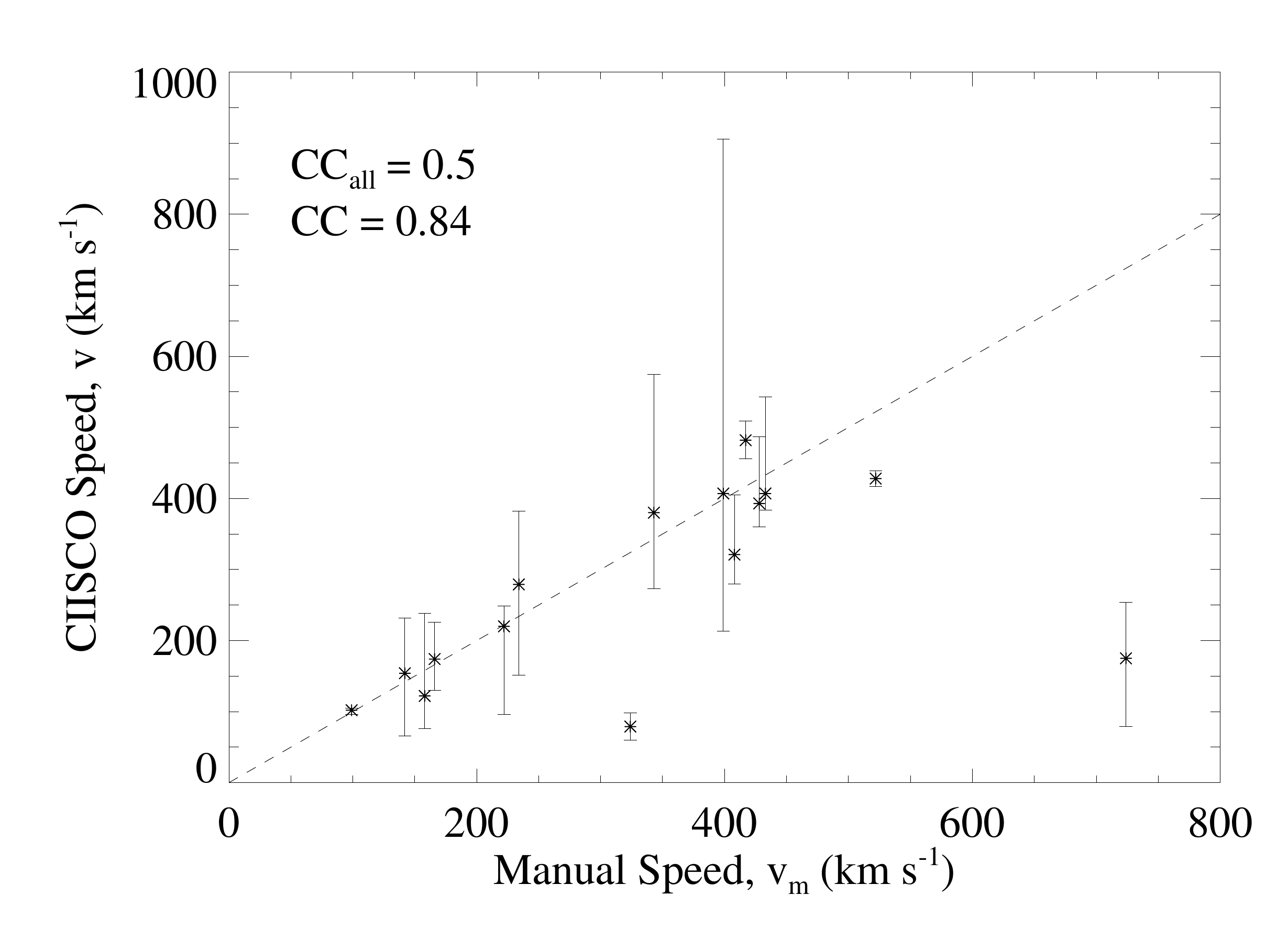}
               \hspace*{-0.002\textwidth}
               \includegraphics[width=0.6\textwidth,clip=]{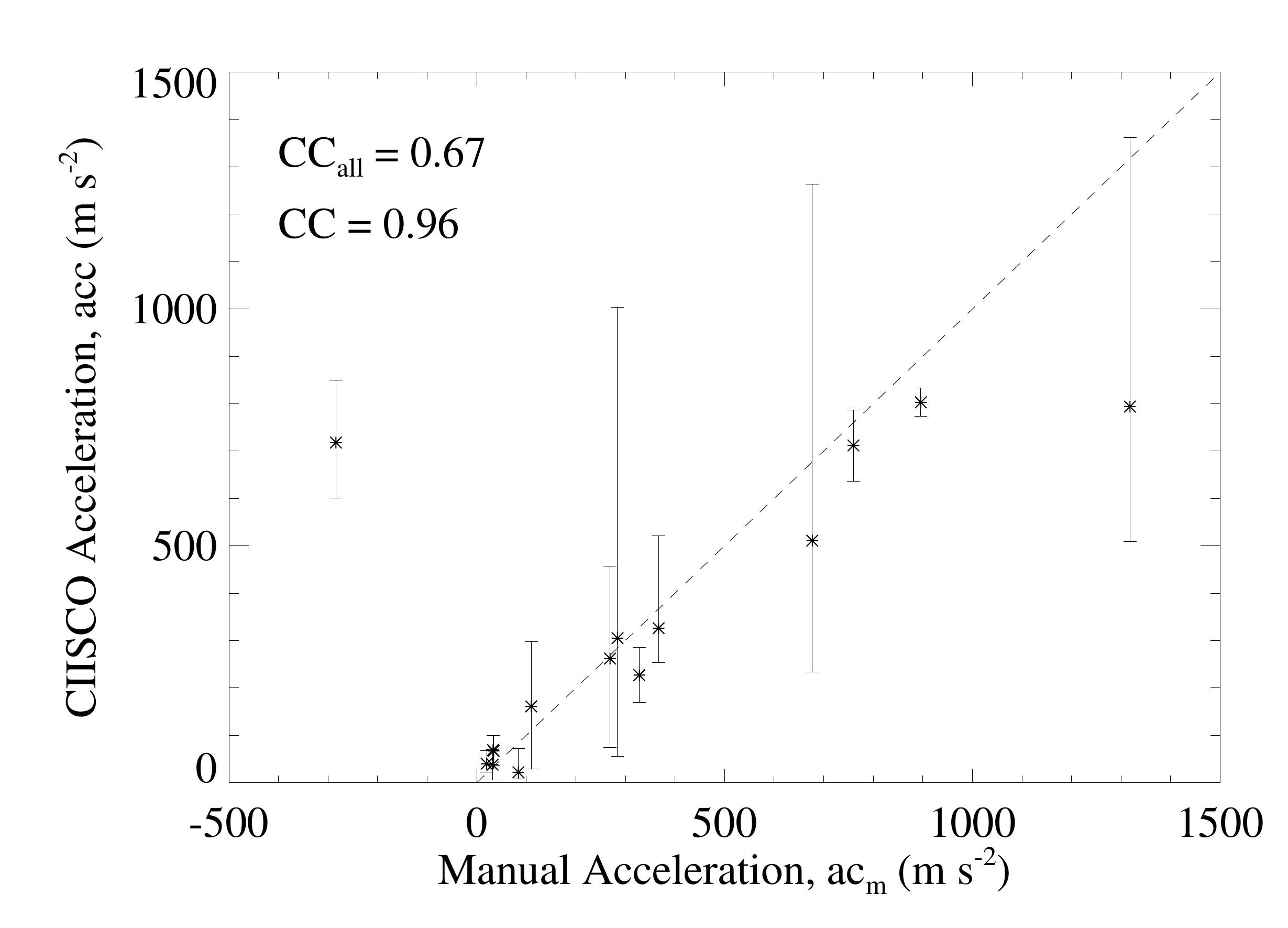}
              }
%
\caption{A comparison plot of the kinematics of eruptions derived manually and CIISCO with speed on the {\it Left} panel and acceleration on the {\it Right}. The dashed line represents y=x. CC$_{all}$ is the correlation value when all the data-points are considered, whereas when one outlying point is omitted the resultant correlation value is given by CC. }
   \label{fig:correlate_plot}
\end{figure}

A closer inspection of Table \ref{table1}, reveals:
\begin{itemize}
     \item CIISCO works well with different EUV datasets. It detected 22 eruptions, of which, 7 were false positive following manual analysis. 
    \item The kinematic properties of the most of the correctly identified solar eruptions derived from CIISCO are close to the values obtained by manual identification. {  Most of these values obtained after manual tracking are within the range of speed and apparent acceleration determined by CIISCO. For such cases, the average speed derived by CIISCO and those computed manually agree within 50 - 100 km s$^{-1}$. {The eruptions 2 and 7 are the prominences erupting from a height $>$1 R$_{\odot}$ and shows deviation in the kinematics properties but are detected adequately.} The value of apparent acceleration after manual analysis comes within the range of values given by CIISCO for 12 out of 15 true detections. For the rest 3 true cases, the difference in the values of apparent accelerations (ac$_m$ from either minac or maxac) is less than 50 m s$^{-2}$.}
    \item { The kinematics derived by CIISCO is compared with manual estimates as shown in Figure \ref{fig:correlate_plot} with the error bars representing the range of values determined by CIISCO. For the speed estimates there is a correlation of 50\% between the two methods when all the data points are considered. This value increases to 84\% when the rightmost value of 724 km s$^{-1}$ is considered as an outlier. The acceleration values shows a good correlation of 67\%. If the one decelerating eruption is left out, then the agreement increases to 96\%.}
    \item CIISCO is able to identify multiple solar eruptions produced at the same location with similar, but different, onset times. The two solar eruptions identified in AIA 304 \AA\ observations on 2012-April-08 at CPA 228$^\circ$ have onset times of 00:15UT and 00:41UT respectively. 
    \item CIISCO also identified solar eruptions at different locations erupting with a time difference of $\approx$ 30 minutes as can be seen for solar eruptions 2 and 3 observed on 2013-May-13 by EUVI-A at 304 \AA\ .
    \item CIISCO wrongly identified a decelerating {  solar eruption} on 2014-July-08 as an accelerating one. This is due to the fact that only ridges defined by Equation \ref{parabola} are identified.
    \item { The eruption of 2014-08-24 was tested in AIA as well as SWAP FOV in similar pass-band. It could be seen that the speed range is similar for the two cases, with the difference of the eruption being accelerating in lower heights.}
\end{itemize}

\setlength{\tabcolsep}{4pt}
\begin{sidewaystable}
\centering
\caption{{Solar eruption} parameters derived from the application of CIISCO algorithm to EUV images 
\label{table1}}
\begin{tabular}{cccccccccccccc}
\hline
Instrument & 
\begin{tabular}[c]{@{}c@{}}Date \\[0.75ex] {\footnotesize(YYYY-mm-dd)} \end{tabular} & 
\begin{tabular}[c]{@{}c@{}}Serial \\[0.75ex] No.\end{tabular} & \begin{tabular}[c]{@{}c@{}}t$_{\circ}$\\[0.75ex](HH:MM)\end{tabular} & \begin{tabular}[c]{@{}c@{}}CPA\\[0.75ex] (Degree)\end{tabular} & \begin{tabular}[c]{@{}c@{}}v \\[0.75ex] \footnotesize(km s$^{-1}$) \end{tabular} & 
\begin{tabular}[c]{@{}c@{}}minv\\[0.75ex] \footnotesize(km s$^{-1}$)\end{tabular} & 
\begin{tabular}[c]{@{}c@{}}maxv\\[0.75ex] \footnotesize(km s$^{-1}$)\end{tabular} & 
\begin{tabular}[c]{@{}c@{}}v$_m$ \\[0.75ex] \footnotesize(km s$^{-1}$) \end{tabular} & 
\begin{tabular}[c]{@{}c@{}}acc \\[0.75ex] \footnotesize(m s$^{-2}$) \end{tabular} & 
\begin{tabular}[c]{@{}c@{}}minac \\[0.75ex] \footnotesize(m s$^{-2}$) \end{tabular} & 
\begin{tabular}[c]{@{}c@{}}maxac \\[0.75ex] \footnotesize(m s$^{-2}$) \end{tabular} & 
\begin{tabular}[c]{@{}c@{}}ac$_m$ \\[0.75ex] \footnotesize(m s$^{-2}$) \end{tabular} & Remarks \\ \hline


\multirow{5}{*}{\begin{tabular}[c]{@{}c@{}}AIA\\[1ex] (171 \AA)\end{tabular}} 
& 2012-04-08 & {  1} & 02:00 & 115 & 1742 & 1361 & 2449 &  &	16836 & 8726 & 29145 &  & False\\
 & 2012-06-27 & {  2} & 09:28 & 315 & 79 & 60 & 98 & 324 &	69 & 37 & 100 & 33  \\
& 2012-06-27 & {  3} & 10:34 & 315 & 279 & 217 &  507 &  & 394 & 219 & 1210 &  & False\\
& 2012-06-27 & {  4} & 11:37 & 73 & 593 & 389 & 773 &  & 1735 & 754 & 2693 &  & False\\
& 2012-08-31 & {  5} & 19:41 & 110 &  428 & 417 & 439 & 522 & 803 & 773 & 833 & 896 \\
& { 2014-08-24} & {  6} & { 12:04} & { 102} &  { 407} & { 384} & { 543} & { 433} & { 794} & { 509} & { 1362} & { 1318} \\
\hline

\multirow{4}{*}{\begin{tabular}[c]{@{}c@{}}AIA\\[1ex] (304 \AA)\end{tabular}} 
& 2012-04-08 & {  7} & 00:15 & 228 & 175 & 79 & 254 & 724  & 161 & 29 & 298 & 110 \\
& 2012-04-08 & {  8} & 00:41 & 228 & 220 & 96 & 249 & 222  & 227 & 169 & 285 & 328\\
& 2012-04-08 & {  9} & 01:24 & 228 & 256  & 244 & 263 &  & 309 & 278 & 324 &  & False\\
& 2014-07-08 & {  10} & 16:15 & 63 & 393 & 360 & 487 & 428 & 718 & 601 & 850 & -284 & Decelerating \\&&&&&&&&&&&&&{eruption}\\
\hline

\multirow{3}{*}{\begin{tabular}[c]{@{}c@{}}EUVI-A\\[1ex] (304 \AA)\end{tabular}} 
 & 2013-05-13 & {  11} & 05:36 & 143 & 174 & 130 & 226 & 166 & 40 & 22 & 68 & 20 \\
& 2013-05-13 & {  12} & 07:56 & 282 & 154 & 66 & 232 & 142 & 38 & 6 & 70 & 32 \\
& 2013-05-13 & {  13} & 08:26 & 81 & 122 & 76 & 238 & 158 & 22 & 8 & 72 & 84 \\ \hline

\multirow{2}{*}{\begin{tabular}[c]{@{}c@{}}EUVI-B\\[1ex] (304 \AA)\end{tabular}} 
& 2012-08-31 & {  14} & 07:16 & 267 & 237 & 171 & 256 &  & 72 & 33 & 148 &  & False\\
& 2012-08-31 & {  15} & 19:26 & 249 & 407 & 213 & 906 & 399 & 305 & 55 & 1004 & 284\\
\hline


\multirow{7}{*}{\begin{tabular}[c]{@{}c@{}}SWAP\\[1ex] (171 \AA)\end{tabular}} 
 & 2011-12-24 & {  16} & 11:28 & 287 & 102 & 96 & 105 & 99 & 67 & 29 & 35 & 34 \\
 & 2012-04-16 & {  17} & 17:57 & 81 & 279 & 151 & 382 & 234 & 262 & 74 & 457 & 269 \\
 & 2012-04-16 & {  18} & 21:04 & 76 & 349 & 301 & 391 &  & 424  & 280 & 547 &   & False\\
& 2013-05-01 & {  19} & 02:23 & 76 & 321 & 280 & 405 & 408 & 326 & 253 & 521 & 367  \\
 & 2013-06-21 & {  20} & 03:04 & 110 & 380 & 273 & 575 & 343 & 511 & 234 & 1264 & 677  \\
 & 2013-06-21 & {  21} & 17:20 & 287 & 586 & 452 & 723 &  & 1188 & 725 & 1857 &   & False\\
 & 2014-08-24 & {  22} & 12:03 & 124 & 482 & 456 & 509 & 417 & 712 & 636 & 787 & 760  \\
\hline

\end{tabular}
\end{sidewaystable}

{We also identified the location of the eruptions that had been correctly identified by CIISCO. We found that eruptions 5 and 10 occured near -65$^{\circ}$ longitude, eruptions 11 and 20 were found to have their origin near -75$^{\circ}$ longitude whereas eruptions 6, 13, 15 and 22 happened at around $\pm$80$^{\circ}$ longitude region. The remaining correctly detected eruptions were the off-disk ones. Thus the eruptions tested here occurred within $\approx$25$^{\circ}$ longitudes from the limb and have been identified as radially moving outward features.  It can be noted from Table \ref{table1} that for eruptions 2 and 7 are the prominence eruptions, which even though occur at the limb, shows some deviation in the kinematics properties. Other than these two, we also found that except for eruptions 5 and 10 which occur near -65$^{\circ}$ longitude, the range of kinematics estimated by CIISCO have values close to manual identification.} { The estimates of kinematics of these eruptions located within 25$^\circ$ from the limb is also complemented with Figure \ref{fig:correlate_plot}. This emphasizes that CIISCO is well suited to be implemented to the eruptions very close to the limb but can provide an approximation of the eruption characteristics for other events like 5 and 10.}

\begin{figure}[!ht]   
   \centerline{\hspace*{0.05\textwidth}
               \includegraphics[width=0.6\textwidth,clip=]{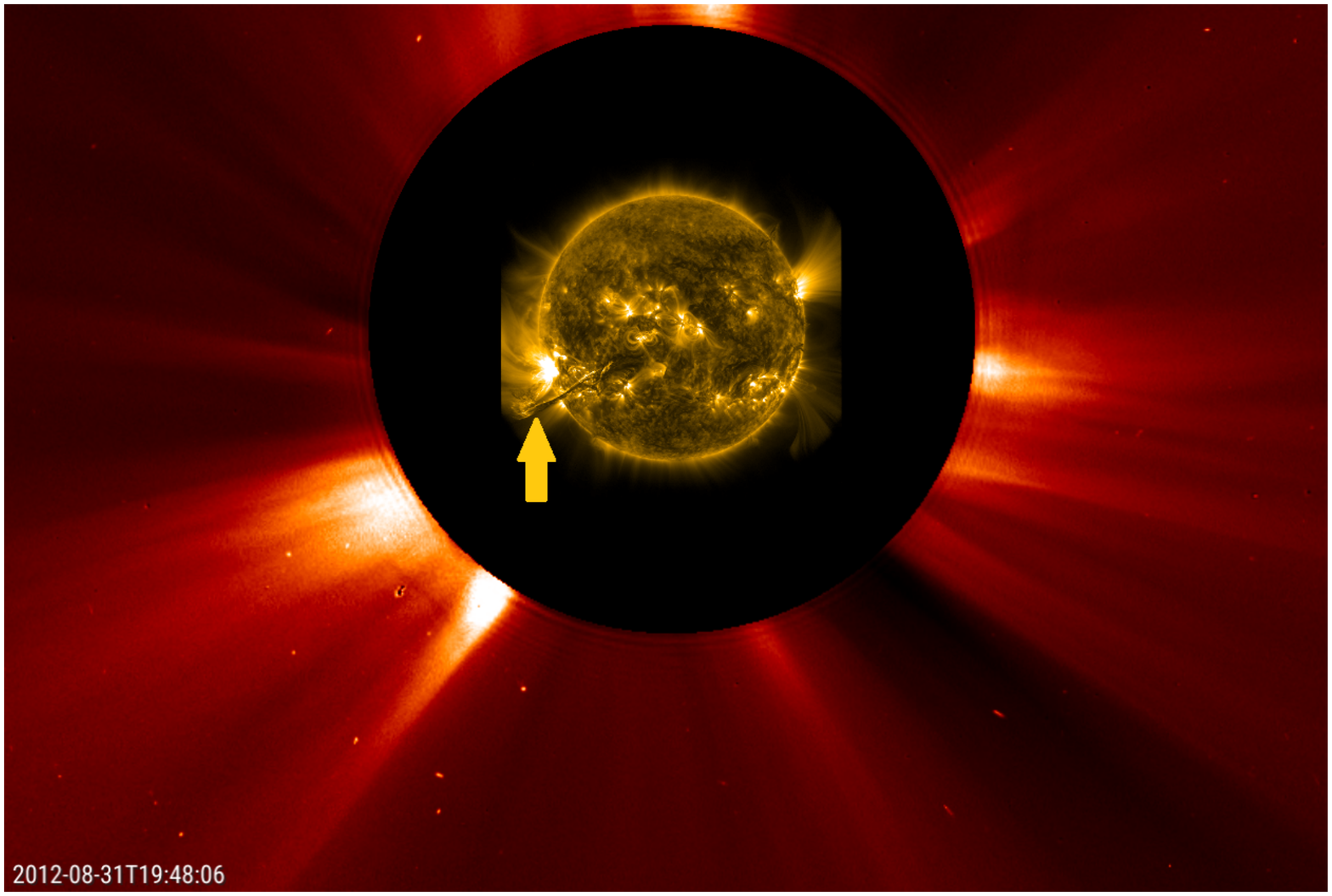}
               \hspace*{0.002\textwidth}
               \includegraphics[width=0.6\textwidth,clip=]{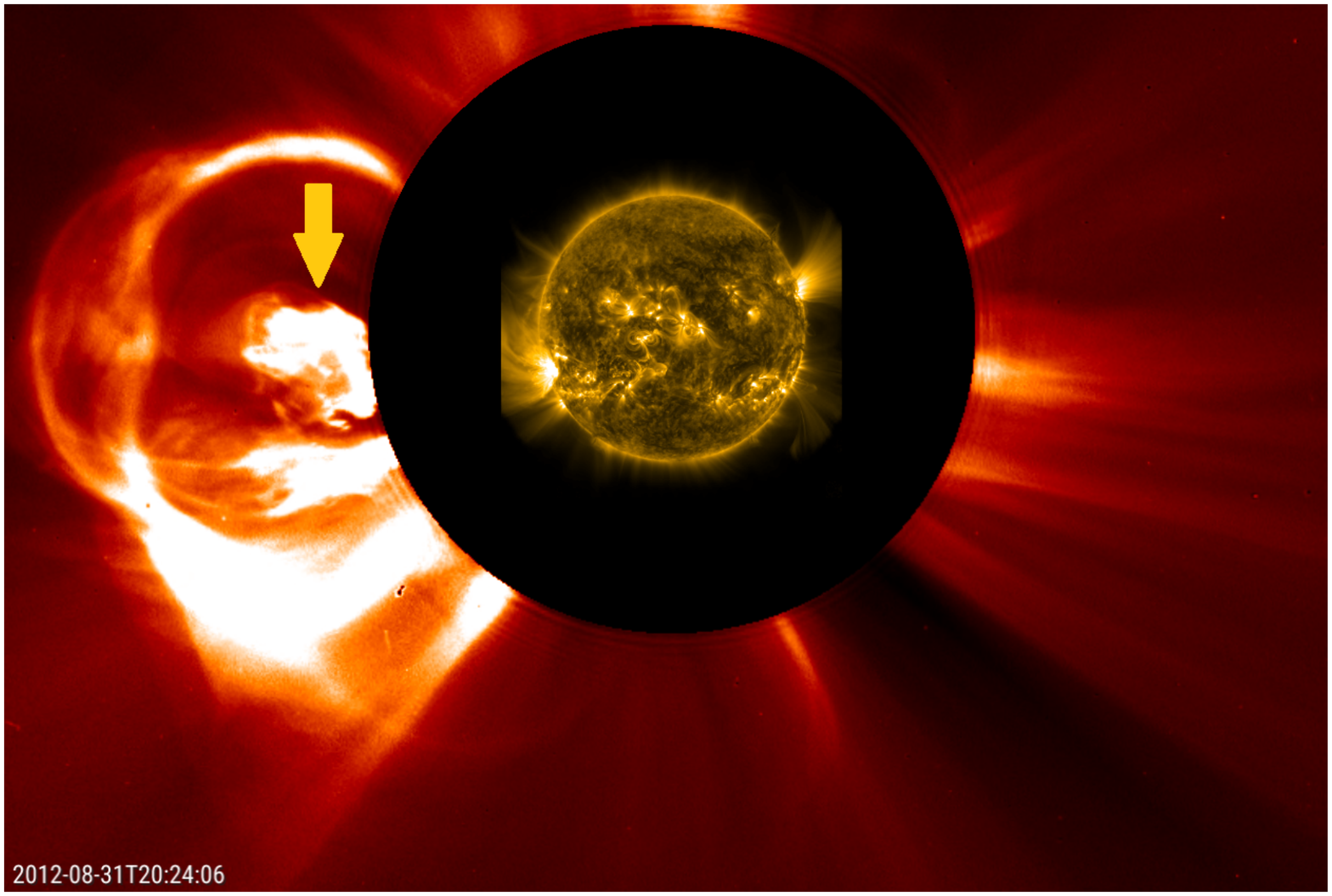}
              }
      \vspace{-0.01\textwidth}  
     \centerline{    
      \hspace{0.175\textwidth}  \color{black}{(a)}
      \hspace{0.6\textwidth}  \color{black}{(b)}
         \hfill}
     \vspace{0.005\textwidth}     
          
 \centerline{\hspace*{0.05\textwidth}
               \includegraphics[width=0.6\textwidth,clip=]{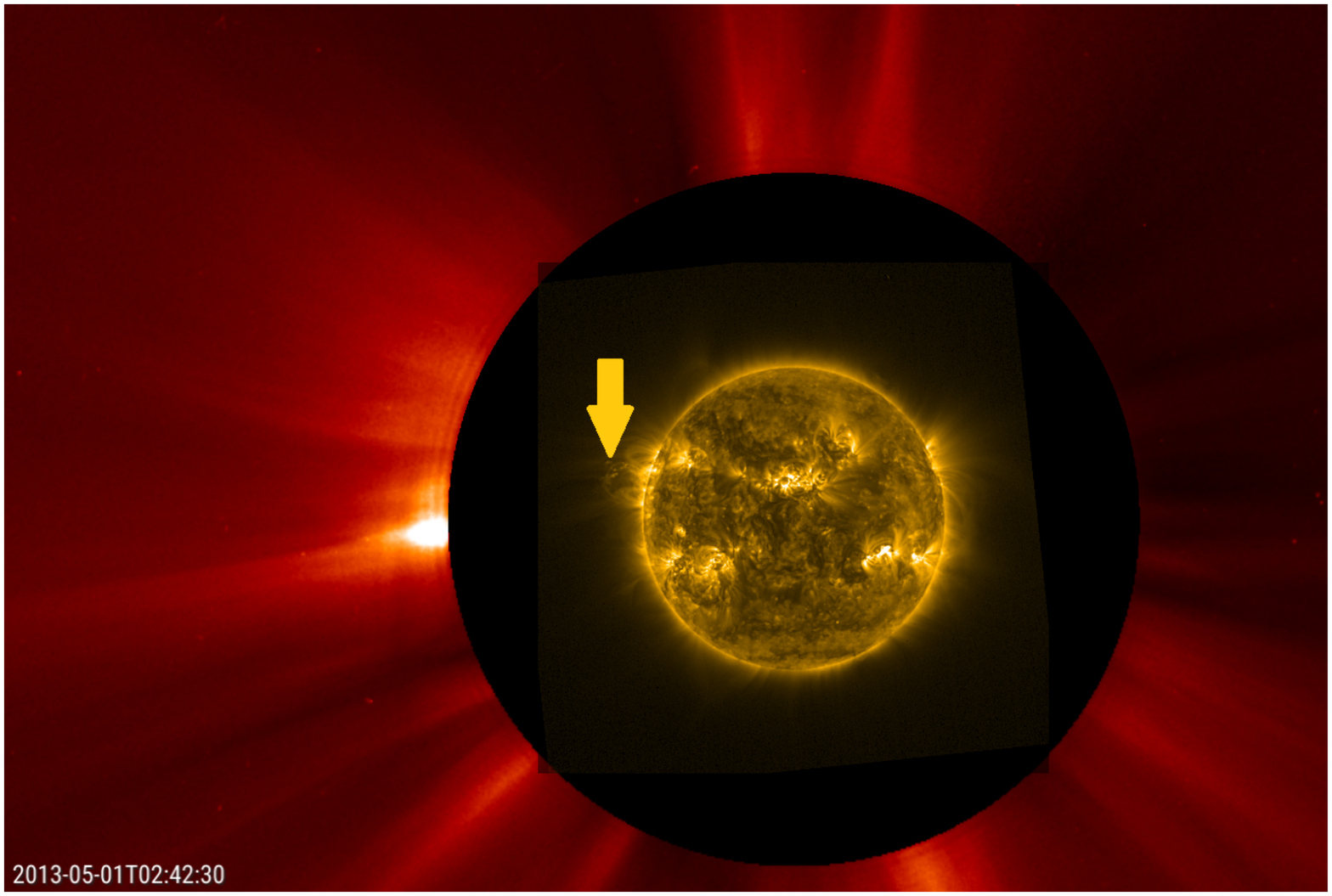}
               \hspace*{0.002\textwidth}
               \includegraphics[width=0.6\textwidth,clip=]{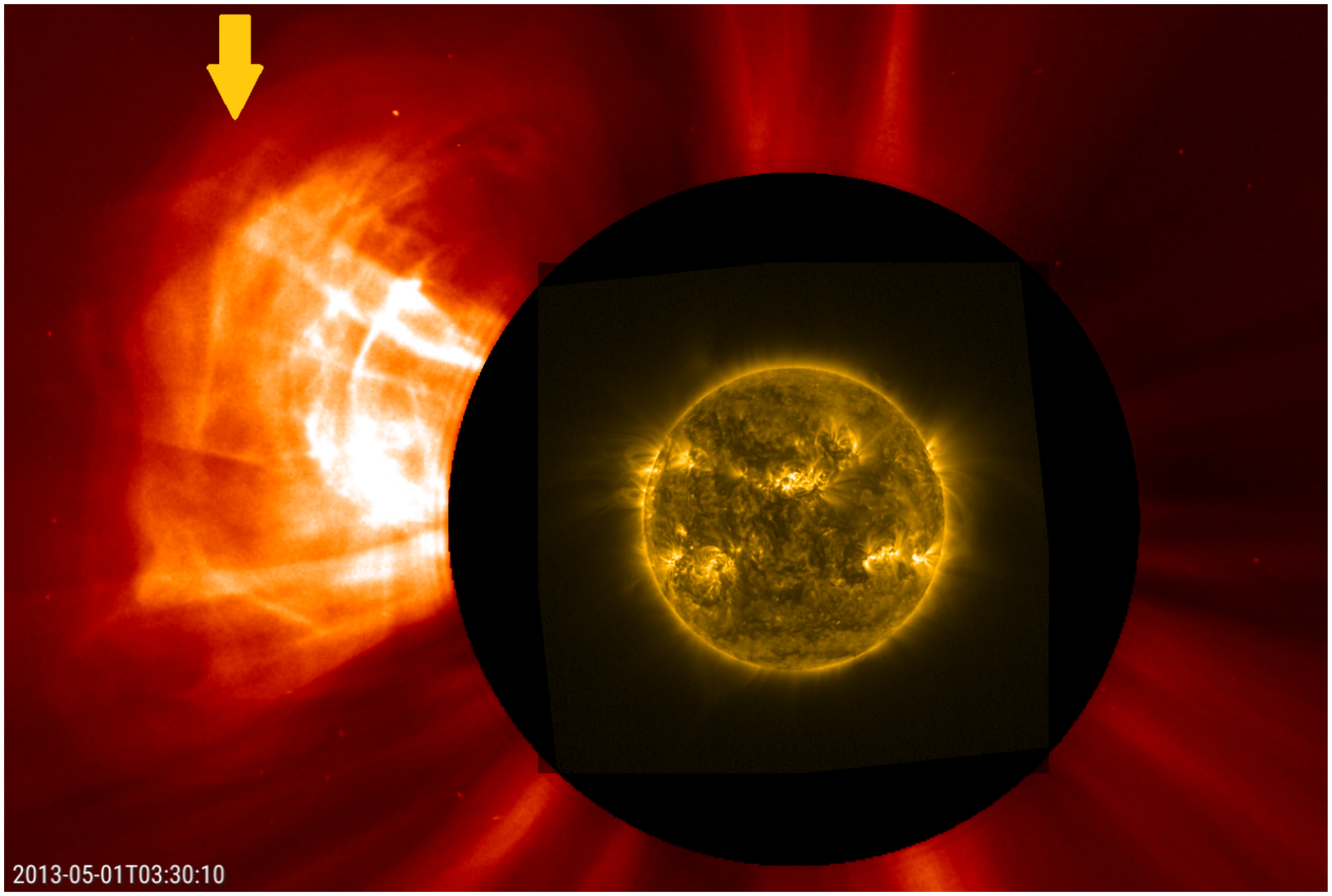}
              }
     \vspace{-0.01\textwidth}  
     \centerline{   
      \hspace{0.175\textwidth}  \color{black}{(c)}
      \hspace{0.6\textwidth}  \color{black}{(d)}
         \hfill}
     \vspace{0.01\textwidth} 

\caption{Solar eruptions as seen in EUV imagers followed in LASCO/C2 coronagraph. {\it Top:} Solar eruption observed on 2012-08-31 at AIA 171 \AA\ that forms CME core as seen in LASCO/C2. {\it Bottom:} Solar eruption observed on 2013-05-01 at SWAP 174 \AA\ that forms CME flux rope as seen in LASCO/C2. The arrow shown in yellow color points to the EUV emission material that propagates and appear as a part of CME in LASCO/C2.}
   \label{fig:counter}
\end{figure}

{  Further for comparison and eruption identification we used the Coordinated Data Analysis Workshops (CDAW) LASCO CME catalog \citep{CDAW2004} and Jhelioviewer \citep{JHelio}. As there is no overlap between the AIA and SWAP FOV and the LASCO/C2 coronagraph these eruptions could not be tracked in the intermediate region and one-to-one comparison could not be made. However we attempted to identify different structures that could be traced in both EUV and white light images.
It turned out that {   eruption 2}, observed by AIA ( 171 \AA\ ), detected on 2012-06-27 and {   eruption 5} (AIA 171 \AA\ ) detected on 2012-08-31 correspond to the cores of the CMEs observed in the LASCO/C2 FOV (see Figure \ref{fig:counter}a,b). {  Eruption 7} (AIA 304 \AA\ ) observed on 2012-04-08 is a prominence eruption that formed the flux rope of the CME seen in the LASCO/C2 coronagraph (see Figure \ref{fig:counter}c,d). However, {   eruption 8} (AIA 304 \AA\ ) on 2012-04-08 is actually part of {  eruption 5}, which split after the prominence eruption and was therefore detected as separate solar eruption. {   Eruption 10} (AIA 304 \AA\ ) on 2014-07-08, {  Eruptions 11 and 12} (EUVI-A 304 \AA\ ) on 2013-05-13, and {  eruptions 16, 17, 19, 20 and 6 and 22} on 2011-12-24, 2012-04-16, 2013-05-01, 2013-06-21 and 2014-08-24 (SWAP) respectively appear to form the flux rope in their corresponding white-light CMEs. {  Eruption 13} observed in EUVI-A 304 \AA\ on 2013-05-13 is a spray\footnote{see \url{https://cdaw.gsfc.nasa.gov/CME_list/autope/}.} type of eruption, often identified as giving rise to narrow CMEs in coronagraph imagery. Figure \ref{fig:splay}(a) shows such an example, where the arrow points to a narrow jet like eruption from an active region that has been identified by CIISCO. Finally, { eruption 15} (EUVI-B 304 \AA\ ) of 2012-08-31 appears to form the core of a CME observed in STEREO/COR1-B FOV.

\begin{figure}[!ht]   
   \centerline{\hspace*{0.05\textwidth}
               \includegraphics[width=0.6\textwidth,clip=]{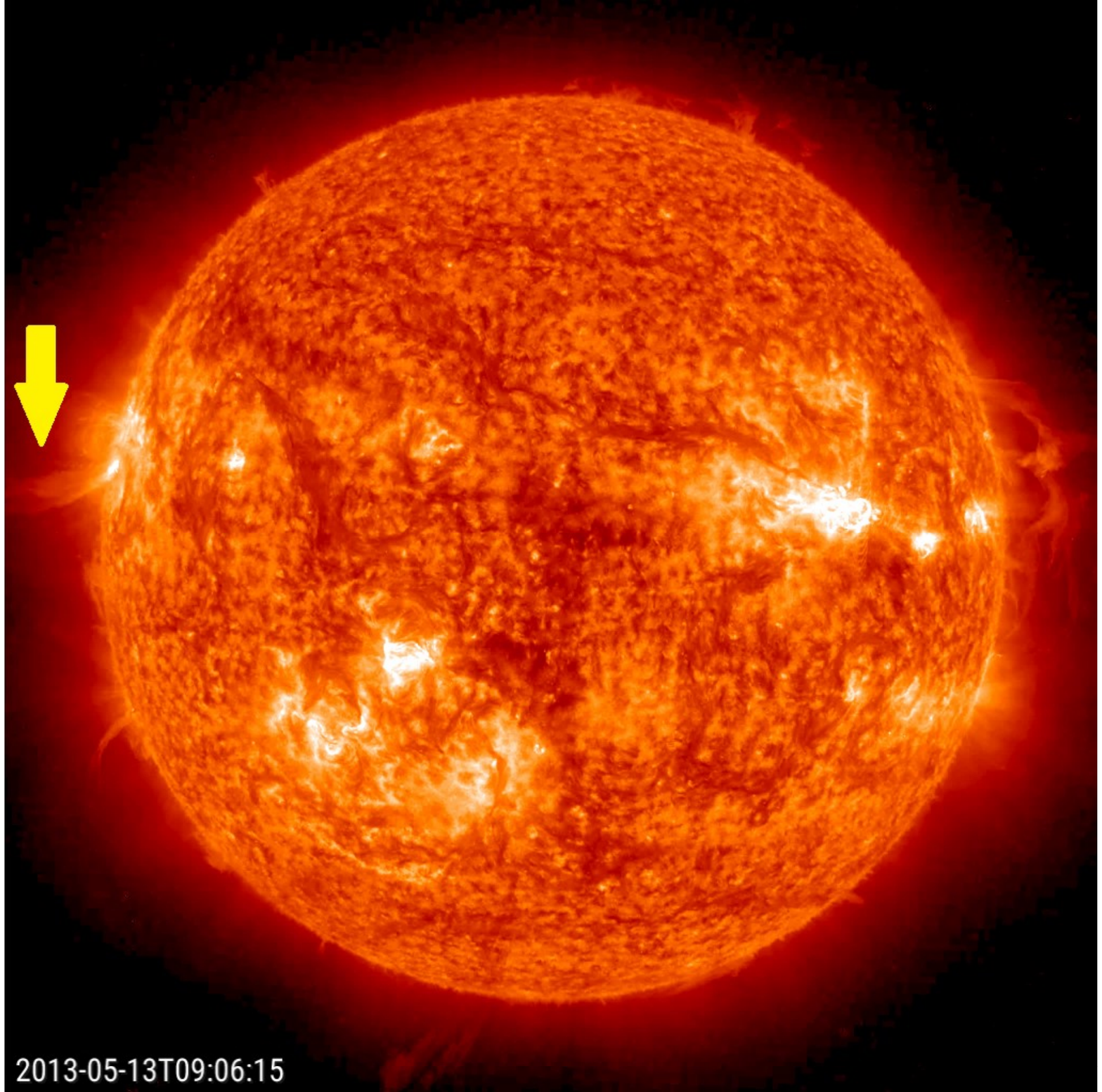}
               \hspace*{0.002\textwidth}
               \includegraphics[width=0.6\textwidth,clip=]{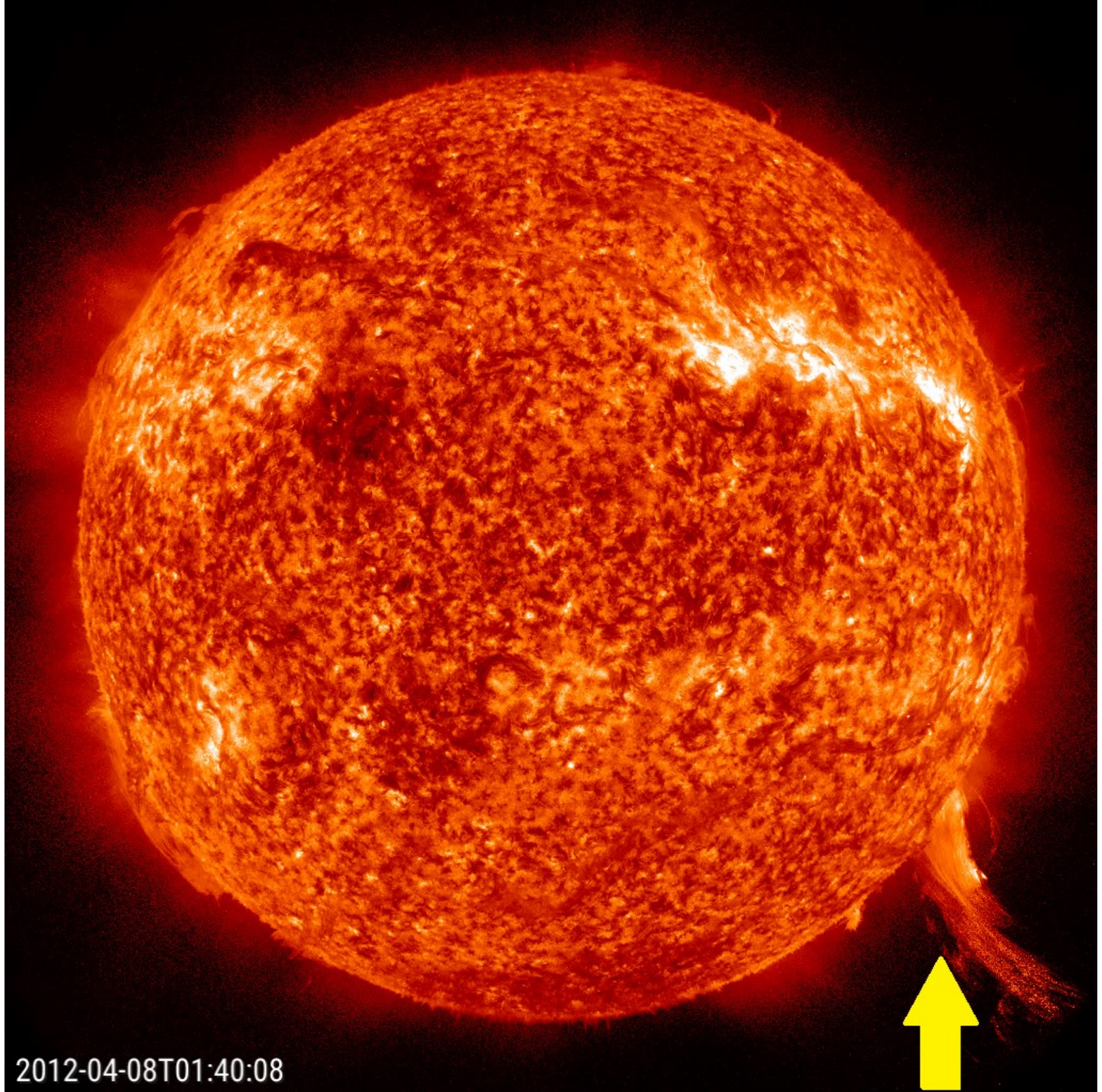}
              }
      \vspace{-0.01\textwidth}  
     \centerline{    
      \hspace{0.175\textwidth}  \color{black}{(a)}
      \hspace{0.6\textwidth}  \color{black}{(b)}
         \hfill}
     \vspace{0.005\textwidth}     
          
\caption{Solar eruptions observed in EUVI-A and AIA. {\it (a)} Solar eruption observed on 2013-05-13 in EUVI-A 304 \AA\ . The eruption pointed by arrow shows a spray type coronal material discharge. {\it (b)} Solar eruption observed on 2012-04-08 in AIA 304 \AA\ where the arrow points to the coronal material tethered to the solar limb after the solar eruption.}
   \label{fig:splay}
\end{figure}

We also found that there are certain cases where CIISCO has detected the coronal material tethered to CMEs creating a false detection of separate CME. These include {  eruption 3} (AIA 171 \AA\ ) observed on 2012-06-27, {  eruption 9} (AIA 304 \AA\ ) on 2012-04-08 and {  eruption 18} observed by SWAP on 2012-04-16. These eruptions have been marked as false positives as they do not correspond to separate outward moving features. An example of such a detection can be seen
in Figure  \ref{fig:splay}(b), where the arrow points to the coronal material bound to the surface. 
{  Eruption 4} (AIA 171 \AA\ ) on 2012-06-27 was also a false detection triggered by the movement in coronal loops. Also, {  eruption 14} (EUVI-A 304 \AA\ ) on 2012-08-31 was a false detection created by a prominence near the western solar limb, solar rotation of this extended feature created a detection recorded as an outward moving structure. This may also be due to the prominence appearing to rise after it had rotated on the limb in subsequent frames. Finally it is noted that an artificial ringing pattern was detected in the threshold of the {  eruption 1}  observed in AIA 171 \AA\ pass-band on 2012-04-08 and {  eruption 21} (SWAP) on 2013-06-21. The faint ringing patterns introduced in the motion filtered images due to Fourier filtering are identified as the primary reason for this, creating false detections. These preliminary results indicate an initial detection efficiency of CIISCO to be
68\%, as 7 out of 22 detections have been identified as false positives.}

\begin{figure}[!ht]   
   \centerline{\hspace*{0.05\textwidth}
               \includegraphics[width=0.6\textwidth,clip=]{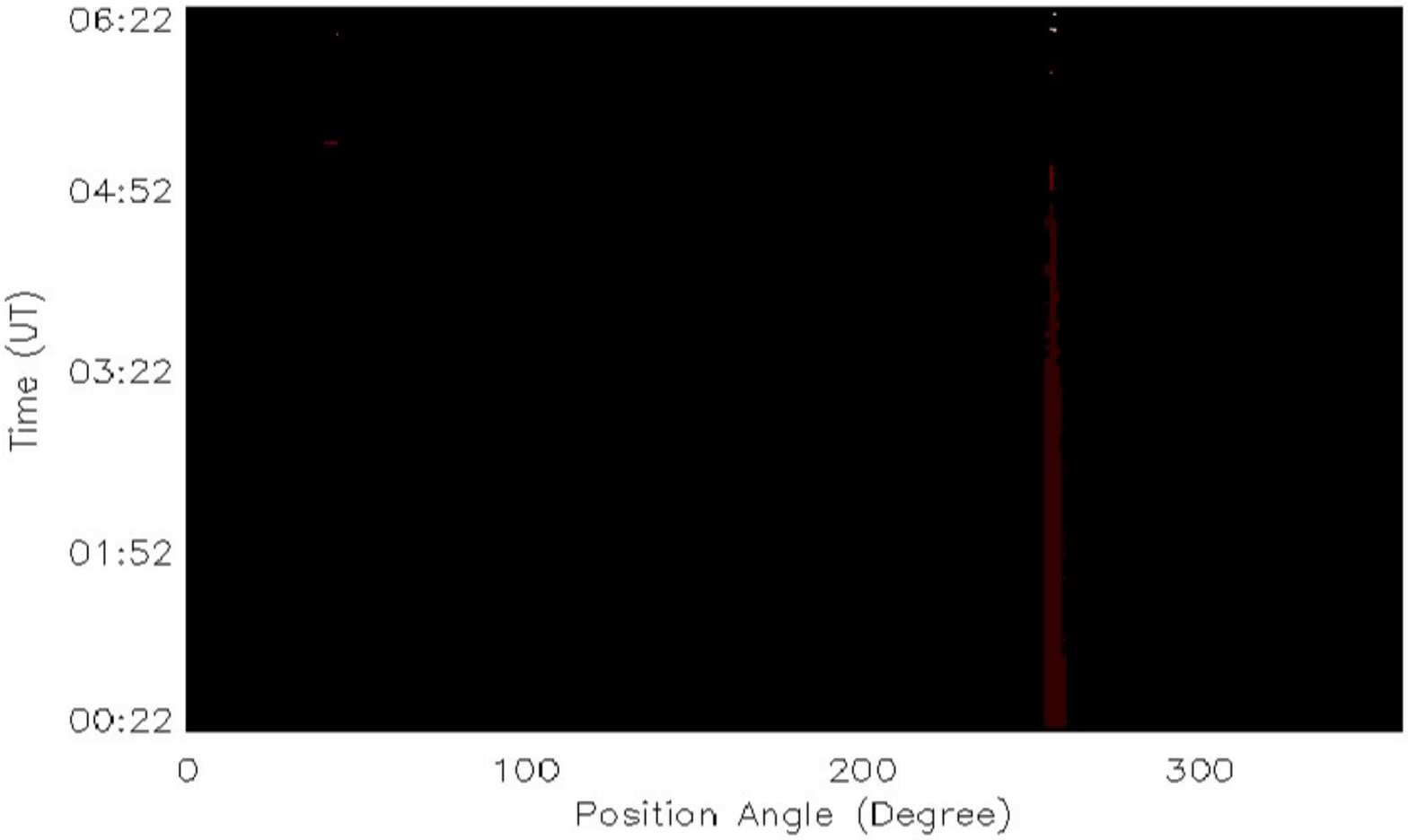}
               \hspace*{0.002\textwidth}
               \includegraphics[width=0.6\textwidth,clip=]{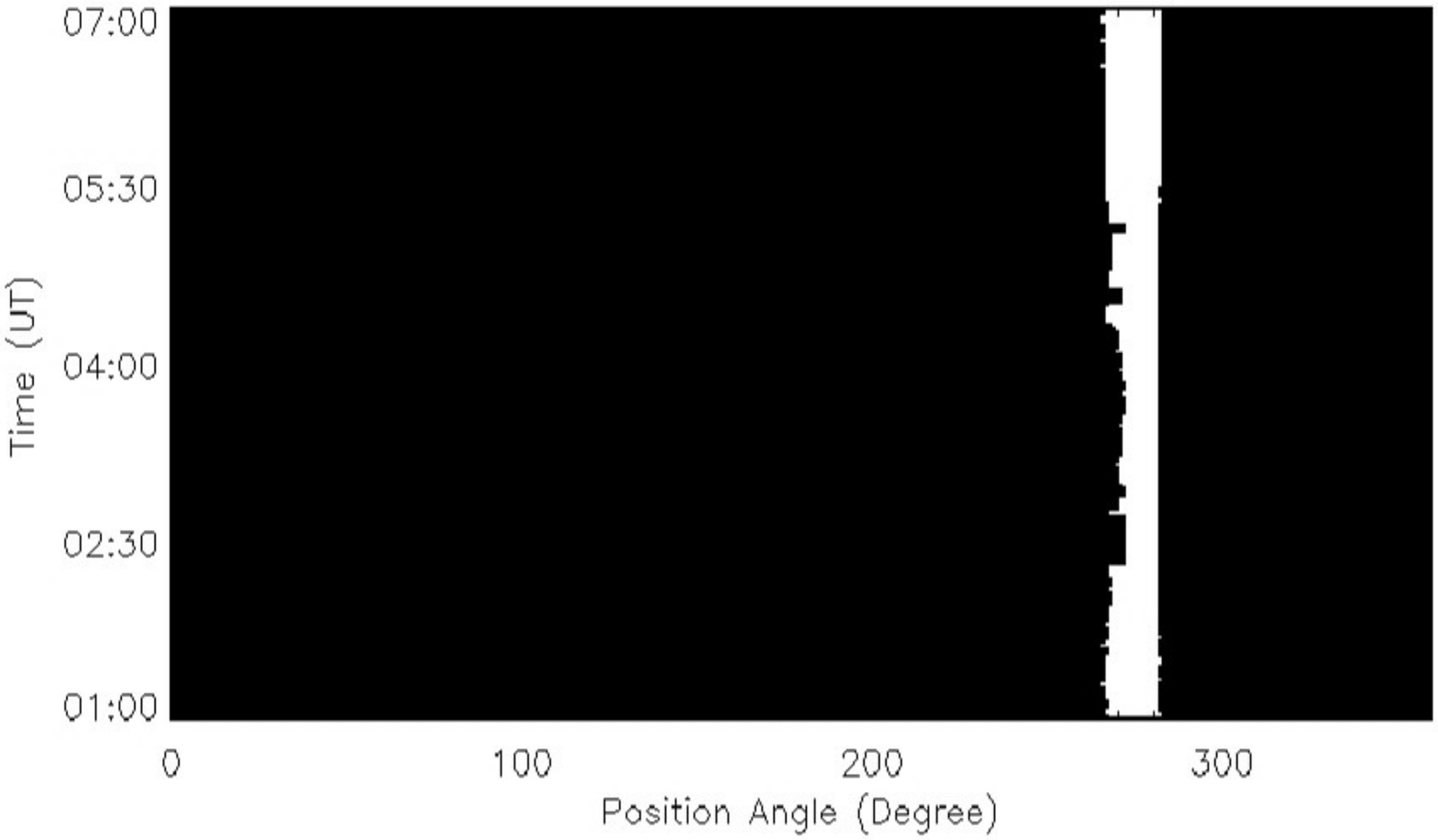}
              }
      \vspace{-0.01\textwidth}  
     \centerline{    
      \hspace{0.175\textwidth}  \color{black}{(a)}
      \hspace{0.6\textwidth}  \color{black}{(b)}
         \hfill}
     \vspace{0.005\textwidth} 
     
   \centerline{\hspace*{0.05\textwidth}
               \includegraphics[width=0.5\textwidth,clip=]{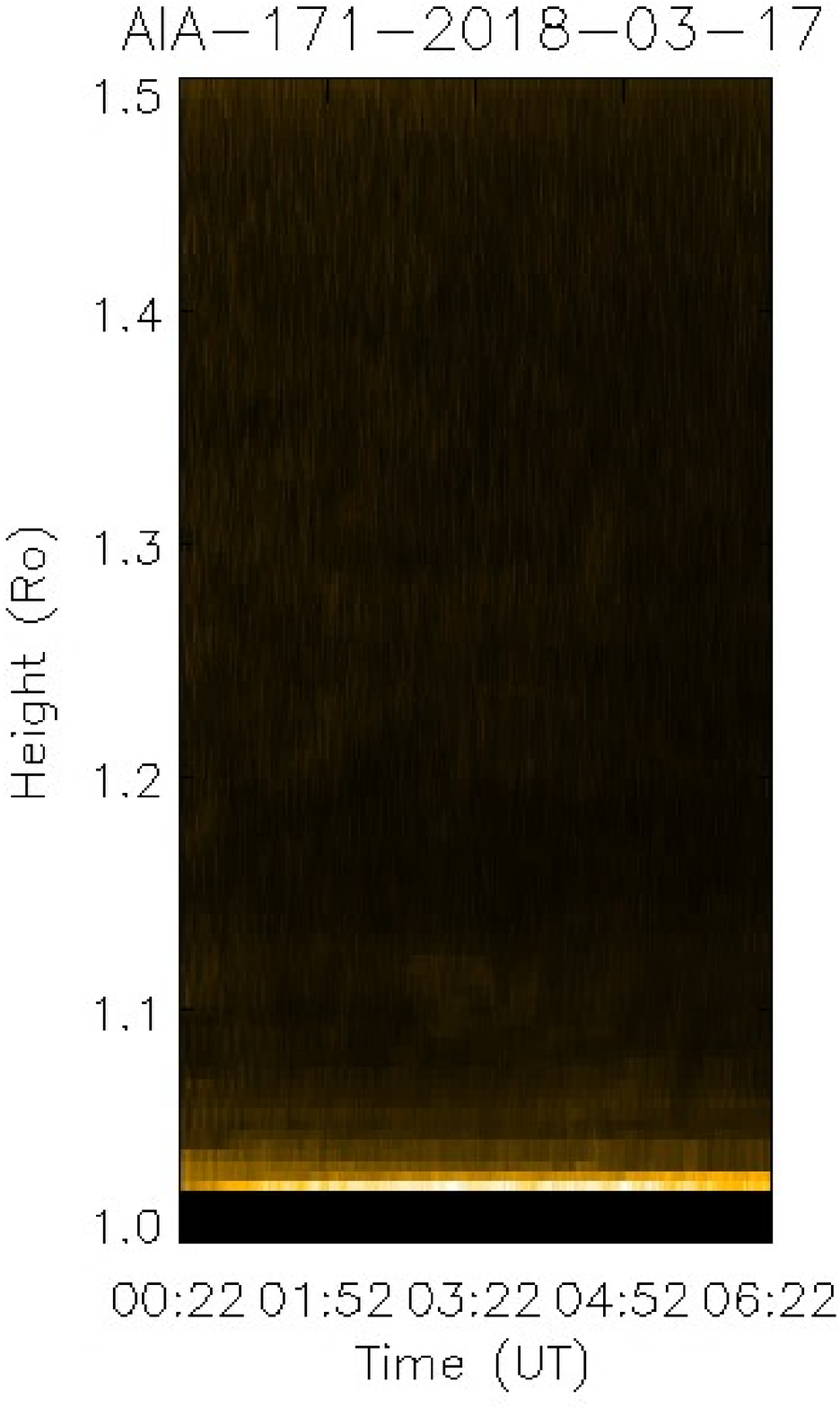}
               \hspace*{0.002\textwidth}
               \includegraphics[width=0.5\textwidth,clip=]{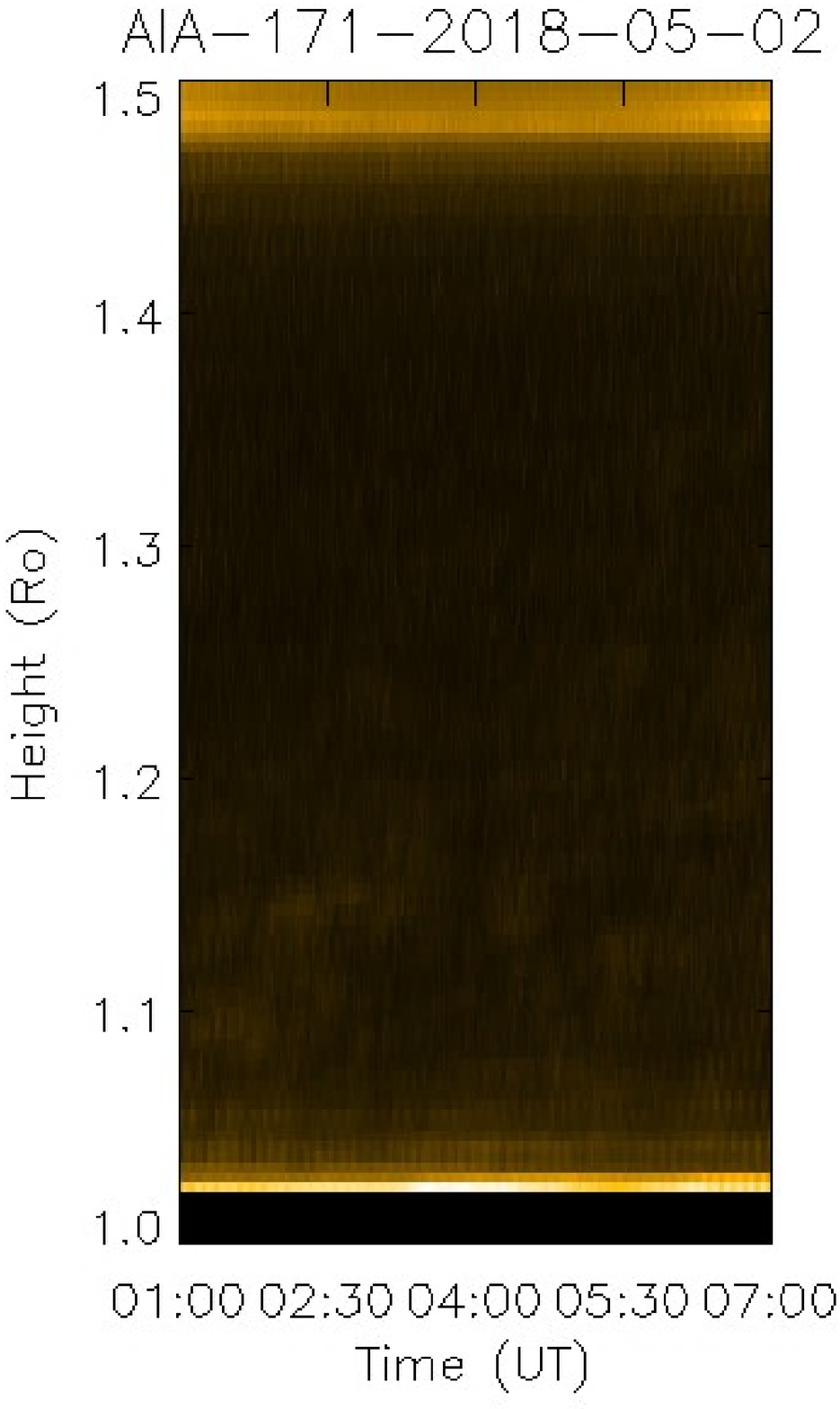}
              }
      \vspace{-0.01\textwidth}  
     \centerline{    
      \hspace{0.275\textwidth}  \color{black}{(c)}
      \hspace{0.5\textwidth}  \color{black}{(d)}
         \hfill}
     \vspace{0.005\textwidth}
          
\caption{CIISCO applied to images of AIA 171 \AA\ pass-band with no solar eruptions. {\it Top:} CME map generated for the test datasets without eruptions. {\it Bottom:} Height-time plots for the two test cases. The left panel shows the output of images taken on 2018-03-17 , whereas the right panel shows output for those taken on 2018-05-02. There are no ridges detected in the two case due to unavailability of any rising feature.}
   \label{fig:nocme}
\end{figure}

{ To determine the false negatives, we looked for eruptions missed by CIISCO during the period of application. We found that for AIA 171 \AA\ observations of the date 2012-04-08, a narrow eruption taking place at $\sim$01:00UT at CPA of $\sim$230$^\circ$ was missed. There was also a narrow jet like eruption of 2014-07-08T17:02 at CPA $\sim$110$^\circ$ observed in 304 \AA\ pass-band of AIA which was missed by CIISCO. These two eruptions being narrow in width may not have satisfied the thresholds and hence unable to be identified. This implies that out of 17 actual eruptions taking place during our period of analysis 15 have been correctly identified by CIISCO giving an efficient detection rate of $\sim$88\%.}
{We have also tested CIISCO for the period when no CME was reported in CDAW and CACTus catalogs. These include periods from 00:00UT to 06:22 UT on 2018-03-17 and similar time period for images observed on 2018-05-02. We used AIA 171 \AA\ images for this test and found that CIISCO does not detect any eruption in the tested time period (Figure \ref{fig:nocme}). For the two test cases, we found that CME regions have been identified in the CME maps as shown in Figure \ref{fig:nocme}(a,b). This is due to the presence of bright pixels consistently after Fourier filtering. It can also be seen from Figure \ref{fig:algofig1}(c, d) that after the removal of loops by Fourier filtering, bright regions are present near the foot-points of these loops. Similar persistent bright regions are also evident just above the masked height from Figures \ref{fig:application} and \ref{fig:nocme}(c, d) showing the r-t plots. As the intensities have been integrated after Fourier filtering to generate the CME map, the intensity threshold applied at this stage are being satisfied to detect CME regions. Nevertheless, the height-time plots generated in the successive steps do not show the presence of bright ridges (Figure \ref{fig:nocme}(c, d)). The absence of any ridge and application of parabolic Hough transform with the set thresholds does not yield any eruption in these two cases. This validates CIISCO to be also effective when there is no observed eruption.}


\section{Conclusion and Discussions} \label{sec:summary}
It has long been known that a CME kinematics varies throughout its propagation in the heliosphere \citep{Byrne12}. An important period in a CMEs evolution is near its onset, in the lower corona. Limited understanding of CME onset and initial progression is reflected in our inability to accurately predict CME arrival times, which currently have an average accuracy of $\pm$ 10 hours \citep{ccmcscore}.
To automatically detect the solar eruptions in the inner corona, we have developed an algorithm, CIISCO, inspired by CACTus and SEEDS, using the parabolic Hough transform assisted by Fourier motion filtering and a SEEDS inspired 1D integrated intensity plot. {Due to the difficulties in making white-light observations of the lower corona, we have used the EUV regime to identify the solar eruptions in this region.} CIISCO automatically identifies and tracks CMEs' {EUV counterparts as a whole unlike other automated CME detection algorithms.} To the best of our knowledge, this is the first successful demonstration of the application of parabolic Hough transform to automatically detect {solar eruptions}. 
We have tested this algorithm on different EUV datasets from AIA, EUVI and SWAP. {The preliminary results show that when accurately detected the average apparent speed and acceleration of {solar eruptions} are in good agreement with the values computed manually with correlation of over 80\%.} CIISCO is also able to identify successive {solar eruptions} produced by the same source, but separated by short periods of time. { In the sample on which CIISCO was tested, it had been able to identify $\sim$88\% of the eruptions correctly.}

{As discussed in previous sections, the solar eruptions observed in EUV will not have a one to one correspondence to those detected white-light, due to the inherent problems with going from the EUV emission regime into a white-light scattered regime. The eruptions observed in the two wavelength passbands is still a matter of debate. Therefore, we also traced the EUV eruptions in to the coronagraph FOV to identify their white-light counterparts. We found that 3 positively detected EUV eruptions correspond to the cores of CMEs observed in white-light, whereas 11 correctly identified eruptions corresponded to CME flux ropes. There was also one case of a spray-type eruption that is observed as a narrow CME in coronagraph observations.}

CIISCO is currently a proof of concept, however a significant fraction of detected `eruptions' are false positives. {  32\% of the detections made by CIISCO have been identified to be false after visual inspection.}
As the algorithms mature this number will improve. {  Some of these false detections such as for the case of the 2012-April-08 detection observed in 304 \AA\ pass-band of AIA, proved to be  post-eruption material tethered to the site of eruption (Figure \ref{fig:splay}(b)). { It needs to be worked out how to reduce such detections as solar eruptions.} Few false detections were a result of artifacts produced from the ringing pattern introduced in images after Fourier filtering. The ringing pattern arising from the filtering limits the application to detect the eruptions with sufficient brightness with-respect-to the background.}
In future, to improve this, we will test different masking techniques ({\it e.g.} the Hanning window) in Fourier space, that will act to preserve the structures and at the same time reduce the ringing patterns. {  Also, a reduction in the artifacts will help us to reduce the intensity threshold which in-turn will allow the algorithm to detect fainter solar eruptions. There are a few cases where the rising coronal material give an impression of outward movement, and therefore created a false detection by CIISCO. Such false positives can be reduced by selecting the range of velocities to be filtered out by the Fourier filtering technique. Such filters can be improved with more CME observations and therefore better CME statistics. Increased CME observations will also provide additional knowledge about the kinematics of these eruptions in the lower corona.
{  We also tested CIISCO for the period when there are no reported CMEs in CDAW and CACTus CME catalogs. We applied CIISCO for the images observed in 171 \AA\ pass-band of AIA on 2018-03-17 and 2018-05-02 for a period of 6 hours. We found that CIISCO successfully did not identify any eruption in the tested time period in spite of the shortcomings mentioned above. This also implies that the chances of false detection in absence of eruptions are very low which can be minimised further after including improvements.}}

{It should be noted that though majority of the eruptions first appear in the FOV with heights in the range of 1-1.025 R$_\odot$ (Figure \ref{fig:hist_erupt}), in the case of prominence eruption of 2012-04-08, the eruption of the prominence initiates at a greater height ($\approx$1.1 R$_\odot$). Nonetheless CIISCO has been able to detect the eruption and provide its approximate properties. It is seen from Table \ref{table1} that the eruptions 2 and 7 have disparity in the kinematics values despite the fact that they occurred at the limb. 
Such cases may need further inspection to understand their behaviour subsequent to eruption. Due to the inherent problems with determining large diffuse structures in EUV observations, CIISCO is probably not suitable for individual case studies that would require individual visual inspection to determine an accurate determination of the CME characteristics. However, CIISCO can certainly provide approximate measurements of the eruption properties allowing the user to determine if an individual event deserves further attention.

Currently we have implemented parabolic Hough transform using 2D parameter space against the 4D space by fixing 2 parameters (r$_{\circ}$=1 R$_{\odot}$, $\theta$=90$^\circ$) out of the 4 free parameters ({\it S}, t$_{\circ}$, r$_{\circ}$,  $\theta$). The difficulty in identifying parameters in 4D parameter space and computation cost led us to fix the two free parameters. The primary goal of this work is to automatically detect solar eruptions in the inner corona which has been attained by fixing the two parameters. Determination of the starting height of such eruptions will be an improvement to the current version of the algorithm and will be a part of future work.}
This may be attained by including techniques like Fast Hough Transform \citep{FHT1986139, FHT1995} in addition to the present method followed. It should also be noted that CIISCO has been applied to identify near-limb solar eruptions only. The eruptions occurring near the disk centre will suffer from lot of uncertainities in measurement of height. Moreover, in EUV wavelengths, these on-disk eruptions may not be observed as distinct features to be automatically identified effectively due to the steep radial gradient in the intensity restricting us to test the algorithm in mostly off-disk eruptions.

Following such improvements, CIISCO will be applied over a larger dataset to generate a catalogue of EUV eruptions and their properties, which will be the first of its kind. We intend to enhance this algorithm for its application to the white-light coronagraph images of STEREO/COR1 and K-Cor. Since, these images suffer from a high amount of noise, no automated algorithm has been successful in automatically detecting CMEs in these datasets. { It should be noted that the types of CME CIISCO can potentially detect is up for debate and can be adjusted as such. The community does not adhere to a standard definition of what a CME is and what characteristics a CME has. Some smaller CMEs may be construed as flows \citep{Robbrecht04} near the Sun. The CACTUS algorithm has certain thresholds in place that broadly differentiate between flows and CMEs. Where flows are classified as suspicious in nature often narrow and of low speed. The level to which a CME becomes a flow can be adjusted by the internal parameters in the algorithm. Similarly, the parameters that govern CIISCO can be adjusted according to the users needs.}

It is worth noting that future solar missions, such as ADITYA-L1 \citep{ADITYA2017}, PROBA-3 \citep{Proba3}, Solar orbiter \citep{solarorbiter}, ${\it etc}$., will have coronagraphs observing the inner corona and full disk UV imagers. This algorithm can be applied to these datasets, and will help add to improve our knowledge of CME evolution following and during their eruption. 
{ The use of CIISCO as a viable Earth orientated eruption forecasting tool relies on observations out of the ecliptic, near 90 degrees to the Sun-Earth line. To date such observations have only been made by NASA’s STEREO \citep{STEREO2008} satellites at varying points in they're orbits. However, CIISCO would be an important tool for ESAs Lagrange mission, a mission being designed to be positioned at the L5 Lagrangian point to specifically monitor space weather from its source on the Sun, through the heliosphere, to the Earth. Onboard Lagrange will be the Lagrange eUv Coronal Imager \citep{LUCI_west}, that is being designed with a wide FOV specifically to detect eruptions, in they’re infancy, close to the Sun-Earth line. From the L5 perspective, LUCI will offer observations from approximately 60 degrees to the Sun-Earth line in near real time. With small changes to the CIISCO algorithm, it could be also be run in semi-real time offering an early warning system for potential Earth-bound eruptions.}
A more generalized form of this algorithm using generalised Hough transform \citep{Ballard81} will be able to identify the changes in profile of CMEs throughout their propagation from the lower corona to the heliosphere and hence improve our current CME propagation models, leading to better space weather forecasts.

\begin{acks}
{  We thank the anonymous referee for the valuable comments.}  This work is supported by PROBA2 Guest Investigator Program.
SWAP is a project of the Centre Spatial de Liege and the Royal Observatory of Belgium funded by the Belgian Federal Science Policy Office (BELSPO). VP is supported by the Spanish Ministerio de Ciencia, Innovaci\'on y Universidades through project PGC2018-102108-B-I00 and FEDER funds. VP was also supported by the GOA-2015-014 (KU Leuven) and the European Research Council (ERC) under the European Union's Horizon 2020 research and innovation programme (grant agreement No 724326). We would like to thank Elke D'Huys and Anshu Singh for their valuable suggestions.
The SECCHI data used here were produced by an international consortium of the Naval Research Laboratory (USA), Lockheed Martin Solar and Astrophysics Lab (USA), NASA Goddard Space Flight Center (USA), Rutherford Appleton Laboratory (UK), University of Birmingham (UK), Max-Planck-Institut for Solar System Research (Germany), Centre Spatiale de Li$\grave{e}$ge (Belgium), Institut d'Optique Th$\acute{e}$orique et Appliqu$\acute{e}$e (France), Institut d'Astrophysique Spatiale (France).
We also acknowledge SDO team to make AIA data available. SDO is a mission for NASA's Living With a Star (LWS) program.
\end{acks}

\vspace{0.25cm}
\footnotesize
{\bf Disclosure of Potential Conflicts of Interest}
The authors declare that they have no conflicts of interest.

\bibliographystyle{spr-mp-sola}
\bibliography{pht_ciisco}  

\IfFileExists{\jobname.bbl}{} {\typeout{}
\typeout{****************************************************}
\typeout{****************************************************}
\typeout{** Please run "bibtex \jobname" to obtain} \typeout{**
the bibliography and then re-run LaTeX} \typeout{** twice to fix
the references !}
\typeout{****************************************************}
\typeout{****************************************************}
\typeout{}}

\end{article} 
\end{document}